\documentclass [11pt, A4]{article}
\usepackage{amsfonts,amssymb,amsmath,amsthm}
\usepackage{graphicx,epstopdf}
\newcommand{\hx}{\hat{x}}
\newcommand{\hX}{\hat{X}}
\newcommand{\hd}{\hat{\del}}
\newcommand{\hr}{\hat{r}}
\newcommand{\hrho}{\hat{\rho}}
\newcommand{\RR}{\mathbb{R}}
\newcommand{\ZZ}{\mathbb{Z}}

\newcommand{\SU}{{SU}}
\newcommand{\su}{{{su}}}
\newcommand{\NN}{\mathbb{N}}
\newcommand{\hp}{\hat{P}}
\newcommand{\la}{\triangleright}
\newcommand{\R}{\mathbb{R}}

\newcommand{\Z}{\mathbb{Z}}
\newcommand{\tens}{\mathop{\otimes}}
\newcommand{\extd}{{\rm d}}
\newcommand{\eps}{{\epsilon}}
\newcommand{\id}{{\rm id}}
\newcommand{\del}{{\partial}}
\newcommand{\CF}{{\mathcal F}}
\newcommand{\CG}{{\mathcal G}}
\newcommand{\CD}{{\mathcal D}}
\newcommand{\DD}{{\mathcal D}}
\newcommand{\CC}{{\mathfrak C}}
\newcommand{\CS}{{\mathcal S}}
\newcommand{\cg}{{\mathfrak g}}
\newcommand{\lp}{{l_p}}

\newcommand{\wE}{\widetilde{E}}
\newcommand{\der}{{\varDelta }}
\newcommand{\Del}{\hat\varDelta}

\def\be{ \begin{equation}}
\def\ee{\end{equation}}
\def\bes{\begin{eqnarray}}
\def\ees{\end{eqnarray}}

\title{Noncommutative Harmonic Analysis, Sampling Theory and the Duflo Map in 2+1 Quantum Gravity}

\author{Laurent Freidel\thanks{email: lfreidel@perimeterinstitute.ca},
 Shahn Majid\thanks{email: s.majid@qmul.ac.uk} \\
\\ \centerline{\footnotesize \it ${}^{*\dagger}$Perimeter Institute for
Theoretical Physics} \\
\centerline{\footnotesize \it 31 Caroline st N, Waterloo, ON,
Canada
N2L 2Y5 }\\
\centerline{\footnotesize \it +} \\ \centerline{\footnotesize \it ${}^\dagger$School of Mathematical Sciences} \\
\centerline{\footnotesize \it Queen Mary, University of London, E1
4NS, UK}}

\date{30 December, 2005 -- revised June 2007}

\begin{document}

\maketitle

\begin{abstract}
We show that the $\star$-product for $U(su_2)$, group Fourier transform and effective action arising in
\cite{EL} in an effective theory for the integer spin Ponzano-Regge quantum
gravity model are compatible with the noncommutative bicovariant
differential calculus, quantum group Fourier transform and noncommutative scalar field theory previously proposed for 2+1 Euclidean
quantum gravity using quantum group methods in  \cite{BatMa}. The two are related by a classicalisation map which we introduce. We show, however, that noncommutative spacetime has a richer structure which already sees the half-integer spin information. We argue that  the anomalous extra `time' dimension seen in the noncommutative geometry should be viewed as the renormalisation group flow visible in the coarse-graining in going from $SU_2$ to $SO_3$. Combining our methods we develop practical tools for noncommutative harmonic analysis for the model including radial quantum delta-functions
and Gaussians, the Duflo map and elements of `noncommutative sampling theory'. This allows us to understand the bandwidth limitation in 2+1 quantum gravity arising from the bounded $SU_2$ momentum and to interpret the Duflo map as noncommutative compression.   Our methods also provide
a generalised twist operator for the $\star$-product.
\end{abstract}

\section{Introduction}

It can be expected that any theory of quantum gravity will
generate noncommutative geometries in its next-to-classical
effective description. While 
3+1 quantum gravity remains little understood,  one can at least
explore this proposal in 2+1 dimensions where quantum gravity with 
point sources is fairly well understood at least in the Euclideanised verison.
It has been known for some decades that such a theory can be cast as a Chern-Simons gauge theory with Lie algebra $e_3$ of Euclidean motions. Briefly, the dreibein collection of three 1-forms is viewed as a gauge field with values in the translation part of $e_3$ while the gravitational connection has values in the rotational part. There is a canonical inner product on $e_3$ needed in the construction of the action. Moreover, as explained in \cite{Wit} such a theory at the classical level is integrable and hence closely tied to quantum groups at the quantum level. Chern-Simons theories are, moverover,  topological and by now well-known to quantise to topological quantum field theories producing knot and 3-manifold invariants controlled by quantum groups. Thus, the Jones knot invariant and related 3-manifold invariant can be constructed directly from the quantum group $U_q(su_2)$, a story now well understood and historically one of the motivations for interest in such quantum groups in the 1990s. 

If one looks in this way at Euclidean 2+1 quantum gravity with positive cosmological constant the corresponding quantum symmetry group is given by the Drinfeld quantum double $D(U_q(su_2))$. The partition function of the theory is then expressed in terms of  the Turaev-Viro invariant \cite{TurVir}.  The limit $q\to1$ is usual Euclidean 2+1 quantum gravity (without cosmological constant); so according to these explanations the latter is controlled by the Drinfeld double quantum group $D(U(su_2))$. Moreover the  partition function is given in this case by the Ponzano-Regge model \cite{PR0}.
The action of this quantum group on the physical states of the theory was explicitly identified in \cite{BaiMul,Sch}. It is only this easiest case which we shall study, leaving  $q$-deformation results for a sequel.

One of the key question of quantum gravity then is to unravel what exactly is the effective geometry description, if any, that stems from integrating out the quantum fluctuation of the geometry.
At least in three dimensions this effective geometry is indeed noncommutative.
The case for noncommutative geometry as an effective theory specifically for Euclidean 2+1 quantum gravity was first proposed by 't Hooft \cite{tHooft}.
It was argued that the noncommutative `coordinate algebra' generated by hermitian operators $\hx^{i}$ is isomorphic to $U(su_2)$
with relations \be\label{su2} {[}\hx^{i},\hx^j{]}= 2\imath\lp
\epsilon^{ij}{}_k\hx^{k}, \ee 
We
denote this non commutative spacetime algebra $\hat{C}_\lp(\R^3)$. As explained in \cite{BatMa},  the quantum double $D(U(su_2))$ indeed acts on $\hat{C}_\lp(\R^3)$ as quantum Euclidean quantum
group of motions deforming $U(e_3)$ with parameter $\lp$. Then (\ref{su2}) should be
viewed as noncommutative version of the `model spacetime' for the theory. A quantum differential calculus, noncommutative plane waves labelled by $SU_2$ momenta and other aspects of the noncommutative geometry were also developed in \cite{BatMa}. The precise physical role of the $\hx_i$ variables was not, however, identified.

Recently in \cite{EL}  this algebra (\ref{su2}) was indeed realised concretely in a construction of Euclidean 2+1 quantum gravity coupled to matter fields.
It was shown that after integrating out the quantum gravity fluctuation one obtains effectively Feynman rules coming from a matter field propagating in the noncommutative spacetime $\hat{C}_\lp(\R^3)$. The approach here was from the Ponzano-Regge state-sum model\cite{PR0} to realise the  quantum gravity path integral. We recall in more detail that  one triangulates the 3-manifold, sums over spins $\{j_e\}$ assigned to each edge $e$ with weight given by 6-j symbols computed over all tetrahedra $t$ formed by nearby edges, i.e. with partition function
\[ Z\sim \sum_{\{j_e\in\Bbb N\}}\prod_{e\in{\rm Edges}}(j_e+1)\prod_{t\in{\rm Tetrahedra}}(-1)^{\sum_{i=1}^6 j_i(t)}\left\{{j_1(t)\ j_2(t)\ j_3(t)\atop j_4(t)\ j_5(t)\ j_6(t)}\right\}.\]
Here the edges around each tetrahedron are numbered $i=1,\cdots,6$ in certain conventions and the 6-j symbol can be written in terms of a sum over deficit angles. In this way the action directly approximates the Einstein-Hilbert action as the triangulation becomes infinitely fine. The spins here should be limited to $j_e\le L$ for some bound $L$ to control infinities. Using the $q$-6-j symbols associated to $U_q(su_2)$ gives  the Turaev-Viro invariant which can then be interpreted as explained above. In this context, the analysis of  \cite{EL} is based on the observation that one can exchange spin summation by group integrals. These integrals can then be interpreted as integrals over a curved momentum space for the particles, when these are present.
For instance one key formula is the identity
\[ \left\{{j_1\ j_2\ j_3\atop j_4\ j_5\ j_6}\right\}^2=\int \prod_{I=1}^4\extd g_I \prod_{I<J} \chi_{j_{\bar e_{IJ}}}(g_I g_J^{-1})\]
where the integral is over the $SU_2$ group manifold with Haar measure, the group variables $g_I$ are assigned to the four vertices of the labelled tetrahedron in a certain convention and $\bar e_{IJ}$ denotes the edge between the two vertices complementary to $I,J$. Finally, $\chi_j$ is the character in the spin $j$ representation. This is the `group field theory' transform that turns the state sum model into a functional integral with group variables. The physical case of interest is of half-integer spin and the group $SU_2$, but the authors restrict to integer spin or in effect the group $SO_3$. This is a technical limitation of the main new ingredient of \cite{EL}, which is to use the group variable $g$ to label ordinary plane waves on $\vec X\in \R^3$ via an expression
\[ e^{{\imath\over 2\lp}{\rm Tr} \vec X\cdot\sigma g}.\]
where $\sigma_i$ are the Pauli matrices.  As we shall explain in detail in Section 2, such waves should be restricted to $g$ in the `upper hemisphere' of $SU_2$, which the authors identify with representatives of $SO_3$. Using such plane waves the `group Fourier transform' of \cite{EL} replaces our group variables above by ordinary  `position' variables $X^i$ on $\R^3$. The group product is equivalent to a certain $\star$-product of such plane waves which in turn defines a $\star$-product for other reasonable functions on $\R^3$. In this way one achieves a $\star$-product quantisation as well as a physical picture of the algebra (\ref{su2}), with  $\lp=4\pi G$ the appropriate Planck length. In a certain `weak gravity' regime it was shown that the effective
Lagrangian now takes the form
\be\label{action1} \int_{\R^3} \del_i \bar\phi\star \del^i\phi\ \extd^3X\ee
in terms of this $\star$-product acting on ordinary scalar fields $\phi(X)$. The $\del_i$ here are the usual partial derivatives. In this way  \cite{EL} showed very
concretely how noncommutative and eventually classical
geometry appears from the combinatorial quantum gravity model,
the emergence of the latter being a fundamental problem for modern approaches to quantum
gravity.

In this paper we are therefore motivated to develop this
connection between quantum gravity and noncommutative geometry 
much further, going beyond the $\star$ product itself
to the deeper noncommutative differential calculus and quantum
group Fourier theory of  plane waves. 
We also develop new mathematical  tools such as noncommutative harmonic analysis and sampling theory  to explore further 
the geometry of a noncommutative  spacetime whose dual momentum space is an homogeneous curved manifold. 
 These techniques play a
crucial role in other noncommutative geometries with curved
momentum space (notably the bicrossproduct spacetime
model \cite{AmeMa}) and should likewise play a physical role in the
2+1 quantum gravity model, which we should like to elucidate. They are
moreover, not limited to integer spin.  We shall see that the two approaches to 
broadly match up and can be combined,  with several fundamental implications of interest. 
That the  above effective action essentially coincides with the action for the
noncommutative scalar wave operator in \cite{BatMa} is then shown
in Section~6. The key is a `classicalisation map' $\phi$ which relates noncommutative plane waves of \cite{BatMa} to group-labelled classical waves of \cite{EL} and which we introduce in Section~2 and study throughout the paper. The
quantum differentials are recalled in Section~3, where we show that
the quantum derivatives $\hat\del_i$ relate to the
usual classical ones precisely by the classicalisation map $\phi$ for the $\star$-product. As we will see this is a key defining property of this classicalisation map.

One important  conclusion coming from our comparison is that while 
the $\star$ product approach so far only sees integer spin information, the
noncommutative geometry is capable of seeing the full theory with half-integer spins, which in turn tells us how the $\star$ product might be be improved. First of all,  the actual structure of the noncommutative differential
geometry of this model in \cite{BatMa} is quite subtle and one deep feature
is that it necessarily has an anomalous  `extra time direction'  $\hat\del_0$ in its tangent space,
as  explored recently in another context in \cite{Ma:time}.  Of
particular interest  then  is the hidden role of this extra $\hat\del_0$
direction in 2+1 quantum gravity, and we shall see for example that it indeed
enters into the formulation of the physical fields in Sections~4,5
even in the integer spin theory.  This $\hat\del_0$ also enters into the twist operator in
Section~7, which addresses a different   problem, namely the
casting of the $\star$ product in the form of a cochain twist
along lines proposed in \cite{BegMa}. We do not construct exactly the
proposed cochain twist but something closely related to it. The second unexpected feature of the noncommutative geometry is, as we show, that it necessarily contains a unique plane wave $\zeta(\hat x)$ of maximum (Planckian) momentum $|\vec k|=\pi/\lp$ corresponding to the group element $-1\in SU_2$, not visible in the classical spacetime. In physics we are familiar with the idea that spacetime rotations of scalar functions see only $SO_3$, one needs fermions to see its universal cover $SU_2$. By contrast scalar (but noncommutative) functions in $\hat{C}_\lp(\R^3)$ already see this covering in form of this plane wave $\zeta$.  We shall find that $\hat \del_0$ and $\zeta$ are intimately related. Motivated by this, we provide (Section~4.3) a first look at how to extend the $\star$-product ideas of \cite{EL} to see the  half-integer spin by means of a new $SU_2$-group valued Fourier transform, and find that this too requires the introduction of an extra variable $T$, related to $\zeta$ and to $\hat\del_0$. Now, one should view the change from half-integer spins to integer spins or from the $SU_2$-momentum to $SO_3$-momentum in our terms, as analogous to `coarse-graining' in lattice quantum gravity, i.e. doubling the effective lattice size, this should be a `renormalisation group step'. While our results are quite specific, they hint at this
important physical conclusion for quantum gravity  that the renormalisation group flow  gets inextricably mixed in with the  spacetime geometry and it is this that appears in the effective theory as an anomalous `extra' dimension' in the noncommutative geometry. 

 Our second main result for physics concerns the implications of bounded momentum. This is one of the key features \cite{BatMa} in the present  noncommutative  model, 
that the momentum space dual to position space (\ref{su2}) is not
only curved but compact, namely $SU_2$ or $SO_3$, in contrast say
to the bicrossproduct model.  To explain these points further and also to provide a template for our noncommutative discussion, let us develop briefly some aspects of `sampling theory' in the classical commutative case where the momentum group is $S^1$. The key ideas are best expressed in terms of the following commutative diagram which compares Fourier transform $\CF$ on $S^1$ with Fourier transform $\CF^\infty$ on $\R$, 
 \begin{eqnarray}C(S^1)& {\buildrel \CF,\cong \over\rightarrow} & C(\Z)\nonumber\\ {\rm extn.} \downarrow\uparrow p& &i\downarrow \uparrow {\rm restr.} \label{S1diag}\\
C'(\R)& {\buildrel \CF^\infty,\cong\over \longrightarrow} & C(\R)\nonumber \end{eqnarray}
where for the purposes of this introduction we denote loosely by $C'(\ )$, $C(\ )$ some appropriate class of complex valued functions or distributions (such as $L^2$ or $l^2$ in the discrete case could be one choice but this is not the only choice of interest) such that the Fourier maps are isomorphisms. The down maps are inclusions and the up maps are surjections, meaning that their composites are projection operators on  $C'(\R)$ and $C(\R)$. Here ``{\rm restr}'' denotes simply restricting a function on position space $\R$ to $\Z\subset\R$ and by the isomorphism this induces a surjection $p$ on the left side. Similarly, ``{\rm extn}'' denotes extending a function on the compact momentum space $S^1$ viewed as a bounded region $[-\pi,\pi]$ by zero outside $[-\pi,\pi]$. This gives a function on momentum space $\R$. It induces an inclusion $i$ on the right. In order to keep the group theory content of relevance to us  we are using $S^1$  not an interval with zero boundary conditions, hence this extension map could be discontinuous, but this need not worry us. The maps $p,i$ can easily be computed as
\[  p(\tilde f)(k)=\sum_{n\in\Z} \tilde f(k+2\pi n),\quad \tilde f\in C'(\R),\quad |k|\le \pi\]
\[ i(f)(X)={\sin(\pi X)\over\pi}\sum_{n\in\Z} (-1)^n{f(n)\over X-n},\quad f\in C(\Z).\]
The part of the diagram involving the perodisation map $p$ is the famous Poisson summation formula. The composite projections 
\[ \tilde\Pi={\rm extn}\circ p,\quad \Pi=i\circ {\rm restr}\]
are related by $\CF^\infty$ and we call them {\em compression maps}. On plane waves one has
\[ \Pi(e^{\imath kX})= e^{\imath [k]X}=e^{\imath kX}\zeta^{-2 n_k};\quad k=[k]+ 2\pi n_k,\quad \zeta=e^{\imath\pi X}\]
where we define the fractional and integer parts of $k$  such that $n_k\in \Z$ and 
\[ [k]\in [-\pi,\pi]\ {\rm if}\  n_k=0,\quad [k]\in ]-\pi,\pi]\ {\rm if }\ n_k>0,\quad [k]\in [-\pi,\pi[\ {\rm if}\ n_k<0.\]
 It is a nice exercise in the theory of hypergeometric functions ${}_2F_1$ to verify this directly from the definition of $i$ given above. The map $\Pi$  compresses any $f\in C(\R)$ to what you get if you sample $f$ on $\Z\subset\R$ and view the result back as a function of $X$ by the map $i$. In momentum space it compresses its spectrum (the support of $\tilde f$) into the region $[-\pi,\pi]$. Note that this story has nothing to do with momentum space being curved, it is a property of it being bounded as this circle example shows (the two are
usually confused in the recent literature).  What is important is that we can identify functions $C(\Z)$ with the image of $\Pi$, i.e. with some subspace functions $\CC(\R)={\rm Image}(\Pi)\subsetneq C(\R)$, say.  Such compressed functions on $\R$ have the property of being determined by their values on the integers. They can be considered as obtained by taking the $S^1$ Fourier transform formula but regarding the conjugate variable as real not integer, or equivalently by the  $\R$ Fourier transform but applied to functions in momentum space with support in the finite bandwidth $[-\pi,\pi]$.  The product of $\CC(\R)$, if it is to coincide with that of $C(\Z)$,  is the product of $C(\R)$ projected back  by $\Pi$. It  corresponds to convolution on $S^1$. This completes our slightly off-beat account of sampling theory but in a form relevant to the paper. 

It is clear that sampling theory  \cite{sampling, Kempf} should also be relevant for quantum gravity.   Our following results can therefore be viewed as first steps in the required `noncommutative sampling theory'. Our first problem to develop this is to know what plays the role of $C(\Z)$. This is not fundamentally a problem for those versed in operator analysis, one can take a Hopf-von Neumann algebra version of $C(SU_2)$ and its dual the Hopf-von Neumann group algebra of $SU_2$, for example. However, this is not what we really need for quantum gravity at its present stage of development, we need practical tools for actual plane waves and Fourier computations. Inspired by the $S^1$ case our strategy is to take the image of the noncommutative version of $\Pi$ as a replacement for the group algebra of $SU_2$. Thus if $\hat{C}_\lp(\R^3)$ stands for a noncommutative version of $C(\R^3)$,  we can define $\CF: C(SU_2)\to \hat{C}_\lp(\R^3)$ as an analogue of the composite Fourier transform  $C(S^1)\to C(\R)$ in the circle case. Its image $\hat{\CC}_\lp(\R^3)\subset\hat{C}_\lp(\R^3)$ with projected product plays the role of $C(\Z)$ and so forth. In physical terms, the Fourier dual of $C(SU_2)$ is defined as the subspace $\hat{\CC}_\lp(\R^3)$ spanned by noncommutative plane waves $e^{\imath \vec k\cdot \hx}$ with bounded $|\vec k|\leq \pi/\lp$. We now come to our first surprise: {\em We shall argue in Section~2 that in fact $\hat{\CC}_\lp(\R^3)=\hat{C}_\lp(\R^3)$} for any minimal completion of the noncommutative polynomial algebra such as to contain plane waves. The fundamental reason is that (as we shall prove)  the element $\zeta(\hat x)$ which plays the role of $\zeta$ in the $S^1$ case now obeys $\zeta^2=1$, due to the topology of the momentum bound and the noncommutativity of the momentum group.  Hence all of the possible noncommutative
plane waves {\em already} have the bounded momentum range, i.e. $\hat{C}_\lp(\R^3)=\hat{\CC}_\lp(\R^3)$ is already compressed. This is physically very important. It says that when the noncommutative theory is related to classical fields on $\R^3$ we should keep in mind that the latter are uncompressed. For example, in the effective action for 2+1 quantum gravity in the $\star$-product form of \cite{EL}, we should not integrate over all classical fields $\phi(X)$  in the effective action but only over compressed classical functions that truly correspond to $\hat{C}_\lp(\R^3)$. 

Since  $\hat{C}_\lp(\R^3)$ is already compressed, the diagram analogous to ({\ref{S1diag}) in which the right hand column would be noncommutative, collapses. However, we can revive the analogy again but this time as a comparison between the quantum Fourier transform $\CF$ and the {\em classical} one $\CF^\infty$ on $\R^3$, 
 \begin{eqnarray}C(SU_2)& {\buildrel \CF,\cong \over\rightarrow} & \hat{\CC}_\lp(\R^3)\nonumber\\ {\rm extn.} \downarrow\uparrow p& &i\downarrow \uparrow \DD \label{SU2diag}\\
C'(\R^3)& {\buildrel \CF^\infty,\cong\over \longrightarrow} & C(\R^3)\nonumber \end{eqnarray}
We do this in Section~5, where we find a natural description of the Duflo quantisation map $\DD$. This is normally studied by mathematicians on polynomials as a deformation construction of the enveloping algebra, but our Fourier methods elevate it to a much wider class of functions including plane waves
\be\label{Duflowave}\DD(e^{\imath \vec k\cdot X})=e^{\imath \vec k\cdot\hx}{\sin(\lp |\vec k|)\over\lp |\vec k|},\quad \vec k\in \R^3.\ee
We see that if we want to view $SU_2$ as bounded momentum, we can consider compression from all classical functions on $\R^3$. The Duflo map both does the compression and quantises the result in one go. These results in the paper suggest a second approach to improving the $\star$ product and  classicalisation of the noncommutative geometry, to be developed further elsewhere.  Whereas $\phi(\zeta)=1$, so does not see the half-integer spin aspect of the quantum spacetime, the map $i$ is singular as   $|k|={\pi\over\lp}$ is approached from below and definitely sees it. 

A final significance of  our results is a complete theory, obtained in Section~5,  for  radial functions $f(\hr)$ in centre of the algebra, where $\hr=\sqrt{\hx\cdot\hx+\lp^2}$. These results are methodological, but if noncommutative geometry on $\hat{C}_\lp(\R^3)$ is to be any use for physical computations we need to be able to work with polar  coordinates, gaussians and spherical waves, and this turns out to be entirely possible. We introduce radial quantum delta functions $\hat\delta_j(\hr)$ as the quantum Fourier transform of characters and prove a `radial sampling theorem' that
\be\label{sample}f(\hr)=\sum_{j\in \NN} f(\lp(j+1))\hat\delta_j(\hr)\ee
for all radial $f\in\hat\CC_\lp(\R^3)$. This is the analogue of our remarks in the circle case and tell us that the noncommutative space has a radial part $C(\mathbb N)$. We will see how the Duflo map exactly implements the compression, while the fact that the noncommutative theory is already compressed   appears radially for example in the identity
\be \sin\left({\frac{\pi\hat r}{\lp}}\right)=0,\ee
which we show holds in $\hat \CC_\lp(\R^3)$.  To round off our noncommutative `harmonic analysis' we obtain and study two noncommutative Gaussians $g_\alpha,f_\alpha$ and their $\star$-product
counterpart given in terms of Bessel functions.   All our results are compatible with the noncommutative differential calculus in the model allowing us to use both integral and differential methods freely. 
On the integration side we define $\int=\sum_j (j+1)\chi_j$ where $\chi_j$ are the traces in the $j+1$-dimensional representation, and similarly $\int_+$ for the $SO_3$ case using even representations only. This integral is  translation invariant, a non trivial and useful fact.   We also provide noncommutative spherical waves $\psi_k(\hat r)$ which diagonalise the noncommutative wave operator defined by $\hat\del^0$. For example, such methods could be used to solve the noncommutative hydrogen atom in a different physical interpretation of the algebra as noncommutative 3-space, which was the interpretation in \cite{Ma:time}.

Finally, let us note that $\hat{C}_\lp(\R^3)$ is also the standard quantisation of
the coadjoint space $su_2^*$ with its Kirillov-Kostant bracket.
Its quotient on setting the quadratic Casimir to a suitable
constant is a matrix algebra viewed as a `fuzzy sphere'\cite{ARS}.
However,  such objects (and matrix methods used for them) are not
relevant to us here. A fuzzy sphere having a fixed radius does not see for example all our results about radial functions. Rather, we work with $\hat{C}_\lp(\R^3)$ with structure induced at the level of polynomials by the Hopf
algebra with coproduct \be\label{cop}\Delta \hx_i=\hx_i\tens 1+ 1\tens
\hx_i \ee and canonical noncommutative differential structure
flowing from this addition law. As explained, by $C(\R^3)$ in this paper we
shall mean a suitably large space of `ordinary' functions of $X^i$
of interest in physics including exponentials, Gaussians etc., and $C'(\R^3)$ a suitable Fourier dual of ordinary functions or distributions in momentum space. By
$\hat{C}_\lp(\R^3)$ we shall mean completion to a deformation of $C(\R^3)$ with
noncommuting $\hat x^i$ as above. We shall argue that any reasonable such completion is already compressed in the sense of coinciding with the image $\hat{\CC}_\lp(\R^3)$ of the quantum group Fourier transform and accordingly, for simplicity, we will later identify the two. 

\section{ Classicalisation of the $\star$-product and the element $\zeta$}\label{starprod}

The noncommutative $\star$-product on $\R^3$ coming out of 2+1
quantum gravity is defined c.f. \cite{EL} to be \be\label{star}
e^{\frac{1}{2\lp}\mathrm{Tr}(X|g_{1}|)}\star
e^{\frac{1}{2\lp}\mathrm{Tr}(X|g_{2}|)}
 = e^{\frac{1}{2\lp}\mathrm{Tr}(X|g_{1}g_{2}|)},
\ee where coordinates $\vec X=\{X^i\}$ on $\R^3$ are viewed as a
$2\times 2$ traceless matrix $X\equiv X^{i}\sigma_{i}$, with
$\sigma_i$ the Pauli matrices
 $\sigma_i\sigma_j = \delta_{ij} +\imath \epsilon_{ijk}\sigma^k$.
  $g$ is an $\SU(2)$ group element represented by
$2\times 2$ unitary matrix and $|g|\equiv
\mathrm{sign}(\mathrm{Tr}(g)) g$ so that $|-g|=|g|$. The
group elements can be concretely written as \be \label{gp} g=
P_0 \id + \imath \lp P^i\sigma_i,\quad P_0^2 +\lp^2 P^iP_i
=1, \ee in which case  $|g|$
is the projection of $g$ on the upper `hemisphere' of $SU_2$ where $P_0\ge 0$.
Indeed the $P_i$ are a coordinate system for $SU_2$ as a 3-sphere
with the unit element at the `north pole'. The plane waves $e^{\frac{1}{2\lp}\mathrm{Tr}(X|g|)}$ appearing in (\ref{star}) has a 2:1 dependence and the `group momenta' are actually being labelled by a
quotient space of  $SU_2$  where $g$ and $-g$ are identified, i.e. by an $SO_3$ group element. Geometrically this  is the identification of the upper half of the 3-sphere with the lower half by the antipodeal
map through the origin. Alternatively, which we shall do, we can always chose a representative for an element of $SO_3$ in the open upper hemisphere. Thus $P_i$ with $P_0=\sqrt{1-\lp^2|\vec P|}>0$ is a coordinate system for most of $SO_3$. We miss in this coordinate system the
2-sphere in $SO_3$ corresponding to the equator.

Therefore in this paper we shall in practice work with the map
\be\label{E} E: SU_2\to C(\R^3),\quad g\mapsto E_g(X)\equiv
e^{\frac{1}{2\lp}\mathrm{Tr}(Xg)}=e^{\imath \vec P\cdot \vec X}.
\ee 
which does not see $P_0$ at all and which is only to used in the upper half of $SU_2$ in defining the $SO_3$ theory by restriction. In terms of these the product $(\ref{star})$ will look a little different, involving now a cocycle,  \be\label{starP} e^{\imath\vec{P}_1\cdot \vec{X}}
\star e^{\imath \vec{P}_2\cdot \vec{X}}= e^{\imath (\vec{P}_1
\oplus \vec{P}_2) \cdot \vec{X}}, \ee where $\cdot$ denotes the 3d
scalar product and \be\label{oplus}\vec{P}_1 \oplus \vec{P}_2 = \epsilon(\vec P_{1},\vec P_{2})
\left( \sqrt{1-\lp^2|\vec
P_2|^2}\,\vec P_1 + \sqrt{1-\lp^2|\vec P_1|^2}\, \vec P_2 - \lp
\vec P_1\times \vec P_2\right), \ee with $\times$  the 3d vector cross
product. The factor  $\epsilon(\vec P_{1},\vec P_{2})=\pm 1$ is the sign of
$\sqrt{1-\lp^2|\vec P_1|}\sqrt{1-\lp^2|\vec P_2|} -\lp^2\vec P_{1}\cdot \vec P_{2}$, it is $1$  if both momenta are close to
zero or one of the momenta is infinitesimal and $-1$ when the
addition of two upper hemisphere vectors ends up in the lower
hemisphere. This factor is a two-cocycle which can be used to
express $SU_{2}$ as a central extension of $SO_{3}$.

{}From (\ref{starP}) we easily find \bes\label{Xstar}
X^i\star e^{\imath\vec{P}\cdot \vec{X}}& = ( X^i P_0+ \lp\epsilon^i{}_{jk}X^jP^k)e^{\imath\vec{P}\cdot \vec{X}}\\
 e^{\imath\vec{P}\cdot \vec{X}}\star X^i &=  e^{\imath\vec{P}\cdot \vec{X}}(X^iP_0  - \lp\epsilon^i{}_{jk}X^jP^k).
\ees

Let us now consider the noncommutative plane waves in
\cite{BatMa}. They are labelled by  group elements $g\in
SU_2$ and provide us with a map \be\label{e} e: SU_2 \longrightarrow \hat{C}_\lp(\R^3), \quad g=e^{\imath{k_i}
\lp\sigma^i} \mapsto e_g \equiv e^{\imath \vec k\cdot \hx} \ee with values in any completion of the polynomial enveloping algebra big enough to include exponentials. On the left hand side we use
the Pauli matrix representation (\ref{gp})  and local
coordinates $k_i$ of $SU_2$. However, we shall see that the above map has a well-defined limit at $-\id$ where the coordinate system breaks down, and is hence defined on all of $SU_2$ (this is not obvious).
These local coordinates $k_i$ are clearly related to our
previous coordinates by \be\label{Pk} \vec P=\frac{\sin{\lp
|\vec k|}}{\lp |\vec k|}\vec k,\quad P_0=\cos{\lp |\vec k|} \ee when
$|\vec k|\in[0,{\frac{\pi}{2\lp}}[$ which covers the upper `hemisphere' 
corresponding to most of $SO_3$. When $k^i$ is restricted to be
such that $|\vec k|\in[0,{\frac{\pi}{\lp}}[$ we cover all of $SU_2$ with
the exception of the one point.

To see that $e_g$ is in fact globally defined, let us consider any element of the form
\[ e_n=e^{\imath{\pi\over\lp}\vec n\cdot \hx},\quad |\vec n|\in \NN.\]
Such an element is central because if $\xi\in \su_2$, then $e_n\xi e_n^{-1}$ is a rotation of $\xi$ by a multiple of $2\pi$ about the $\vec n$ axis, i.e. gives back $\xi$. Moreover, for any $k_i$ we have
\[ e^{\imath \vec k\cdot\hx}e_ne^{-\imath \vec k\cdot\hx}=e_n\]
since $e_n$ is central, but the left hand side is $e_{n'}$ for some rotated vector of the same length (and every vector of the same length can be obtained in this way). Hence the elements $e_n$ depend only on the nonnegative integer which is the length of $\vec n$.   Therefore we see that in any reasonable completion of the enveloping algebra we will have an element 
\be\label{zeta} \zeta=e^{\imath { \pi\over\lp} \vec n\cdot \hx}, \quad |\vec n|=1,\quad \zeta^2=1\ee
in the centre, which is the unique value of $e_g$ as $g\to -\id$. The third equality is because we can rotate any $\vec n$ to $-\vec n$ so that $\zeta=\zeta^{-1}$.
We also see that the other elements $e_n$ mentioned above are just powers of this one. The noncommutative plane waves
$e^{\imath k\cdot \hx}$ exist similarly for any Lie algebra $\cg$
 of dimension $n$ say, in the role of $su_2$ above, and live in the corresponding $\hat{C}_\lp(\R^n)$ as a quantisation
  of $\cg^*$ by its enveloping algebra. 
  
  The  quantum group Fourier transform \cite{Ma:book} adapts in this context to
\be\label{qgfou}\CF:C(G)\to \hat{C}_\lp(\R^n),\quad \CF(\tilde f)=\int_{G}\extd g
\tilde f(g)e_g\ee where $G$ has Lie algebra $\cg$ of dimension $n$. The Haar measure here
should be converted to suitable $k^i$ local coordinates and there will be similar issues as above.
{A priori,} the  image of $\CF$ is does not look like it should be all of $C_\lp(\R^n)$  and in general it will not
be, since it is spanned only by (noncommutative) plane waves with bounded
 $\vec k$ according to the range of the coordinates as we go over $G$. In this paper we denote it by $\hat{\CC}_\lp(\R^3)$ in the case of   $SU_2$ and by  $\hat{\CC}_\lp^+(\R^3)$ when we restrict to $g\in SO_3$ by which we mean the upper `hemisphere' in $SU_2$.
 The ordinary plane waves $E_g$ likewise have bounded momentum $\vec P$ and we denote the subalgebra with $P_0>0$ that they similarly generate  $\CC_\lp^+(\R^3)$.

As to the existence of a suitable `completion' to use of the polynomial algebra 
$C^{\rm poly}_{l_{p}}(\R^3)\equiv U(su_{2})$,  we first  introduce a norm on this 
 by using the fact that $SU_{2}$ acts naturally on it by
rotation of the coordinates $\hx^{i}$ and this action $\hx \to
g\la \hx$ preserves the star structure $\hx^\dagger= \hx$ and the
commutation relations. Given this action we define \be |\hat{f}|^2
= \int_{SU_{2}} dg\, g\la(\hat{f}^\dagger \hat{f}). \ee This is an
invariant polynomial function which  belongs to the centre of
$U(su_{2})$. Lets consider now an highest weight (non necessarily
integral) vector $v_{\lambda}$ and $V_{\lambda }$ the
corresponding Verma module. Since $|\hat{f}|^2$ is central its
action on $v_{\lambda}$ is diagonal and the proportionality
coefficient is  denoted $|\hat{f}|^2(\lambda)$. This provide us
with a family of norms. A powerseries $\hat{f}$ is said to be
convergent with radius of convergence $R$ if
$|\hat{f}|^2(\lambda)< \infty$ for all $\lambda<R$. Note that if
$\hat{f}$ belongs to the centre of $U(su_{2})$ it is a function of
the quadratic Casimir $c=\hx_{i}\hx^{i}$,  $\hat{f}\equiv f(c)$ and
then $|\hat{f}|^2(\lambda)= f((\lambda+1)^{2}-1)$. We can now
define $ \hat{C}_\lp^\infty(\R^3)$ to be the space of
powerseries which have a non zero radius of convergence as the 
basis for one possible `completion'.

We now  consider, where defined (which includes near the identity
$g=1$), the map \be\label{phi} \phi: \hat\CC^+_\lp(\RR^3) \rightarrow
\CC_\lp^+(\RR^3),\quad e_g  \mapsto E_g. \ee Under the quantum group
Fourier transform the product of  $\hat{\CC}_\lp(\R^3)$ is equivalent to
the convolution $\bullet$-product on $C(SU_2)$ and
$\delta_{g_1}\bullet\delta_{g_2}=\delta_{g_1g_2}$. We see that the
quantum-gravity $\star$-product is precisely isomorphic to this
convolution product under the composition
$\phi\CF(\delta_g)=\phi(e_g)=E_g$, at least when restricted to
$SO_3$, i.e. the left cell of the following diagram commutes with the
$\bullet$-product and $\star$-product structures on the respective
linear spaces \be
 \begin{array}{lcccr} C(SO_3),\bullet & {\buildrel \CF\over\longrightarrow} & \hat\CC^+_\lp(\R^3)&\hookrightarrow& \hat{C}_\lp(\R^3)\\
&\searrow & \downarrow\phi& & \downarrow\phi \\
& & \CC_\lp^+(\R^3),\star&\hookrightarrow
&C(\R^3),\star\end{array}\label{fou}\ee The composite $\phi\CF$ is
the map called the `group Fourier transform' in \cite{EL} and we
see that it connects via $\phi$ to the previous quantum group
Fourier transform \cite{Ma:book}. 

{}From our definition it is clear that the map $\phi$ is an
isomorphism in the middle position and in principle could be expected to extend as depicted 
on the right to a generic isomorphism between
$\hat{C}_{\lp}(\R^3)$ and $C(\R^3)$ at least at the level of polynomials
and hence formally to functions given by power series. Indeed,
$\phi^{-1}(E_{g})=e_{g}$ and we can think of $E_{g}(\vec X)$ as a
generating functional in the sense that if $F(\vec X)$ is a polynomial
function it can be obtained by repeated derivation
$F(\vec X)=F(\frac{\partial}{\partial P})e^{\imath\vec{P}\cdot
\vec{X}}|_{\vec{P}=0}$. Since $P(k)$ is invertible around $P=k=0$,
$\phi^{{-1}}$ can be defined on all polynomial functions. $\phi$
is also an isomorphism of algebras at this level, i.e. let
$f_{1}(\hx), f_{2}(\hx)$ be two noncommutative functions in
$\hat{C}_\lp(\RR^3)$ then \be \label{hom} \phi(f_{1} f_{2})(X) =
\left(\phi(f_{1}) \star \phi(f_{2})\right)(X). \ee

After some algebra the map $\phi$ can be explicitly written in
terms of the monomials: \be \phi(\hx^{\{i_{1}}\cdots
\hx^{i_{s}\}}) = \sum_{n\in I_{s}}\lp^{s-n} C_{n,s}
\delta^{\{i_{1}i_{2}}\cdots \delta^{i_{s-n-1}i_{s-n}}
X^{i_{s-n+1}}\cdots X^{i_{s}\}} \ee where we have  introduced the
index space $I_{s}=\{n\in\NN/ s-n \in 2\NN\}$, and  the bracket
denotes the symmetrisation of indices. The coefficients $C_{n,s}$
are  given by \be C_{n,s}= \frac{1}{2^{n}}\sum_{k=0}^n (-1)^k
\frac{(n-2k)^{s}}{(n-k)!(k!)}. \ee These coefficients also satisfy
\bes
C_{n,s}&=& \frac{s!}{n!}\sum_{k_{1}+\cdots +k_{n}=\frac{s-n}{2}} \frac{1}{(2k_{1}+1)!\cdots (2k_{n}+1)!}\\
&=& \frac{1}{2^{n}n!} \sum_{\epsilon_{i}=\pm
1}\left(\prod_{i=1}^n\epsilon_{i}\right)(\epsilon_{1}+\cdots
+\epsilon_{n})^s \ees and \be \frac{1}{n!} \left(\frac{\sin
\theta}{\theta}\right)^n = \sum_p C_{n,n+2p}
\frac{(-1)^p\theta^{2p}}{(n+2p)!}. \ee They are such that $C_{s,s}
=1$, so the image of a given monomial contains this monomial plus
lower order terms obtained by
 Wick contraction $\langle \hx^{i}\hx^j\rangle = \lp \delta^{ij}$.

There could still be specific functions on which $\phi$ does not extend or extends but is not injective. In particular, we have
 \be\label{phizeta}
\phi(\zeta)=1=\phi(1)\ee
because the evaluation of the left hand side is then equivalent to computing $\phi(e_g)$ as $g$ approaches the south pole, which
will converge and be given by $E_{-\id}=1$. This is exactly why we restricted to the patch to $P_0>0$ in the first place for $SO_3$
and this issue essentially limits the amount of information in $\hat C_\lp(\R^3)$ that $\phi$ can see. 
We shall see  momentarily that essentially $\hat \CC_\lp(\R^3)=\hat C_\lp(\R^3)$ and shall see later that the former is
an extension of  $\hat\CC_\lp^+(\R^3)$ precisely by $\zeta$. We conclude that the extended $\phi$ still has its image in $\hat\CC^+_\lp(\R^3)$,  that
(\ref{phizeta}) is the only source of degeneracy and that the extension in the right hand column of  (\ref{fou})  is
just the trivial one by $\phi(\zeta)=1$. In short, the $\star$-product and its classicalisation
map $\phi$ can only ever see the $\hat C^+_\lp(\R^3)\subset \hat C_\lp(\R^3)$ information coming from $SO_3$.

Note also that to have such an isomorphism between $\hat{C}^{\rm poly}_\lp(\R^3)=U(su_2)$ and
polynomial functions on $\RR^{3}$ with some $\star$ product and to
extend it to formal powerseries is not unusual, the most common
example is to take the inverse of the symmetrisation map \be
\sigma^{-1}(\hx^{\{i_{1}}\cdots \hx^{i_{s}\}}) = X^{i_{1}}\cdots
X^{i_{s}}, \ee which defines a $\star$-product on $\RR^3$
\cite{Gut,KL}. In the latter case we would have
$\sigma^{-1}(e^{\imath \vec k\cdot\hat x})=e^{\imath \vec k\cdot
\vec X}$ and $e^{\imath \vec k_1\cdot\vec X}\star e^{\imath \vec
k_2\cdot\vec X} =e^{\imath B(\vec k_1,\vec k_2)\cdot \vec X}$
where $B(\vec k_1,\vec k_2)$ is given by the
Baker-Campbell-Haussdorf formula. However, such a $\star$-product
is not compatible in the same way as above with the quantum group
Fourier transform as the quantum gravity induced one (\ref{star}).
This compatibility property uniquely determines $\phi$ with the features just described.

Finally, let us give an elementary argument that in the case of $SU_2$ momentum group, one has essentially $\hat\CC_\lp(\R^3)=\hat
C_\lp(\R^3)$ for any reasonable coadjoint quantisation. For any $\vec k\in \R^3$ write $\vec k\equiv k \hat k$ where $\hat k$
is a unit vector in the upper half sphere and $k\in \R$. We let $\vec k=[\vec k]+{2\pi\over \lp}n_k\hat k$ where $n_k$ is an integer and $|[\vec k]|\le {\pi\over\lp}$ in similar conventions to those for $S^1$ in Section~1. We have
\be\label{ek}
e_{\vec k}\equiv e^{\imath\vec  k\cdot \hx}=e^{\imath [\vec k]\cdot\hx}e^{\imath {2\pi\over \lp} n_k\hat k\cdot\hx}=e^{\imath [\vec
k]\cdot\hx}\zeta^{2n_k}=e^{\imath [\vec k]\cdot\hx}=e_{[\vec k]},
\ee
where we have used that $\zeta^2=1$.
Since $\hat C_\lp(\R^3)$ is supposed to be a reasonable deformation of usual functions $C(\R^3)$ we suppose (or one can
take this this as a definition) that every element of it  has a Fourier expansion in terms of noncommutative plane waves $e^{\imath \vec k\cdot \hx}$ in analogy with the usual Fourier expansion of $C(\R^3)$. We might naively write it, on the assumption that $\hat{C}_\lp(\R^3)$ has the same `size' as classically, as an integral over $\vec k\in \R^3$ with some Fourier coefficients to be determined. However,  in  view of (\ref{ek}) it suffices to make any such expansion as an integral only over the ball $|\vec k|\le \pi/\lp$, 
\[ f(\hx)=\int_{|\vec k|\le {\pi\over\lp}}\extd^3k\  g(\vec k) e^{\imath \vec k\cdot\hx}\]
for some classical function $g(\vec k)$. 
But the quantum group Fourier transform for $SU_3$ explicitly takes the form of such an integral 
\begin{equation}\label{qgft}\CF(\tilde f)(\hx)=\int_{|\vec k|\le {\pi\over\lp}}\extd^3k \left({\sin(\lp |\vec k|)\over\lp |\vec k|}\right)^2 \tilde f(\vec k)  e^{\imath \vec k\cdot\hx}\end{equation}
and since the Jacobian factor here is nonzero in the interior of the ball, we conclude that any $f(\hx)$ in $\hat{C}_\lp(\R^3)$ can equally well be expressed as a quantum Fourier transform, or arbitrarily-well  approximated as such. We shall understand this more formally in Section~5.3 in terms of the Duflo map.

\section{Quantum differential calculus}

We now show that, among all the possible $\star$-products on
$\RR^3$ with algebra relations $\hat{C}_\lp(\R^3)$, only the one given
in (\ref{star}) is compatible with the bicovariant noncommutative
differential calculus.

Indeed, as for any algebra, one has on $\hat{C}_\lp(\R^3)$ the abstract
notion of a noncommutative differential calculus
$(\Omega^1,\extd)$ where the space of `1-forms' $\Omega^1$ is a
bimodule over   $\hat{C}_\lp(\R^3)$ (it means one can associatively
multiply 1-forms by functions in $\hat{C}_\lp(\R^3)$  from the left and
right), and
\[ \extd :\hat{C}_\lp(\R^3)\to \Omega^1\]
 obeys the Leibniz rule. One also requires that 1-forms of the form
 $f(\hat x)\extd g(\hat x)$ span $\Omega^1$ and that the only thing killed by
 $\extd$ is a multiple of the constant function 1. In our case we have an addition law
 (\ref{cop}) and we require that the calculus is translation-invariant with respect to this.
 The smallest such calculus was found in \cite{BatMa} to be 4-dimensional with basis
 $\extd \hat x^i,\theta$ over the algebra and relations
\be\label{calc} (\extd \hat x^i) \hat x^j-\hat x^j \extd \hat
x^i=\imath\lp\eps^{ij}{}_k\extd \hat
x^k+\imath\lp\delta^{ij}\theta \ee
 \be   \hat x^i \theta - \theta \hat x^i = \imath{\lp}
\extd \hat x^i.\ee The 1-form $\theta$ has no classical analogue
(there is an anomaly for differentiation on quantisation)  and we
are free to change its normalisation; we have used a natural one.
It has the key property \be\label{theta} [f,\theta]=\imath\lp\extd
f\ee on any function $f(\hat x)$. One may identify the space of
left-invariant 1-forms abstractly as $2\times 2$ hermitian
matrices with  $\extd \hat x^i=\lp \sigma_i$ and $\theta=\lp\id$.
Given the choice of basis of $\Omega^1$ the corresponding quantum
partial derivative operators $\hat\del_i,\hat\del_0$ on
$\hat{C}_\lp(\R^3)$ are canonically defined by \be\label{extd} \extd
f=\sum_i(\hat\del_if)\extd\hat x^i+(\hat\del_0f)\theta.\ee On
noncommutative plane waves $e^{\imath k\cdot \hat x}$ the
$\hat\del^i,\hat\del_0$ are computed \cite{BatMa} as
\be\label{partial} \hat\del_i=\imath { \sin({\lp}|\vec k|)\over\lp
|\vec k|}k_i \ee \be\label{del0}
\hat\del_0={\imath\over\lp}\left(\cos(\lp|\vec k|)-1\right)\ee The
computations here do not assume any bounds on $|\vec k|$ i.e.
actually hold in $\hat{C}_\lp(\R^3)$ which is the setting for all the
computations in \cite{BatMa, Ma:time}. The second equation means
that $\hat\del_0$ can be built from
$\hat\nabla^2=\hat\del_i\hat\del^i$ in the physical momentum range
as \be\label{del0lap}
\hat\del_0=\frac{\imath}{\lp}\left(\sqrt{1+\lp^2\hat\nabla^2}-1\right).\ee
We will often use the combination
\be\label{hatDel}\Del\equiv 1-\imath\lp\hat\del_0=\sqrt{1+\lp^2\hat\nabla^2}\ee
where the second expression holds in the physical momentum range.

Comparing (\ref{partial})  with (\ref{phi}) we see immediately
that  $\phi$ intertwines the quantum partial derivatives
$\hat\del_i$ with the usual partial derivatives \be \label{der}
\phi(\hat{\partial}_{i}f)(X) = \partial_{i} \phi(f)(X) \ee since
the latter on plane waves  bring down the $\imath k^i \sin(\lp
|\vec k|)/\lp|\vec k|$ factor exactly as required for
(\ref{partial}). This is therefore a second key property which
distinguishes our homomorphism $\phi$ and associated $\star$
product. As required for consistency with (\ref{phizeta}) we see from (\ref{partial}) that
\[ \hat\del_i\zeta=0.\]
Note, however, that 
\[\hat\del_0(\zeta^2)=0,\quad  \hat\del_0\zeta=-{2\imath\over\lp}\zeta,\quad \Del\zeta=-\zeta\]
where the first is needed for consistency with (\ref{zeta}). The other (equivalent) identities tell us that $\hat\del_0$ `sees' the extension $\hat\CC^+_\lp(\R^3)\subset\hat\CC_\lp(\R^3)$ by $\zeta$ i.e. that this extra tangent direction is linked to the `coarse-graining' from $SO_3$ to $SU_2$. 

It is not surprising that the compatibility with bidifferential
calculus or quantum Fourier transform leads to the same
homomorphism since they are are deeply related. Indeed,  the
partial derivatives $\del_i$ for any bicovariant calculus on a
Hopf algebra is given \cite{Wor} by evaluation of the coproduct
against some elements $\imath P_i$ (say) in the dual, i.e. by the
coregular representation. The characteristic property of the
quantum group  Fourier transform \cite{Ma:book,KemMa} is that
 it intertwines this representation with multiplication in the dual. Thus in general
\be \hat{\partial}_{i} \CF(\tilde f) = \CF(\imath P_{i} \tilde f ), \ee where $\tilde f\in
C(G)$ and the product in the RHS is the commutative product
of functions. The general form of bicovariant calculus for enveloping algebras $U(\cg)$ is described explicitly in \cite{Ma:twi}.

\section{Quantum integration and Fourier theory}

As well as quantum  differential calculus we have integration. In
fact there is no problem to define a translation-invariant
integration on $\hat{\CC}_\lp(\R^3)$ in so far as the results converge:
in view of our result (\ref{der}) whereby $\phi$ intertwines the
classical and quantum differentials, we just define
\be\label{intstar}\int_\phi f(\hat
x)=\int_{\R^3}\phi(f)(X)\extd^3X\ee in terms of usual Lebesque
integration.  This is then translation invariant in the sense \be
\label{intdif}\int\hat\del_i=0\ee which is the sense needed in
practice. Having this for all quantum differentials is more or
less equivalent to translation covariance with respect to the
additive coproduct (\ref{cop}) because the partial derivatives of
a translation-invariant calculus on (any) Hopf algebra are given
by evaluation against the coproduct as explained above. Such
methods have been used routinely on other linear quantum
spaces \cite{Ma:book}.

We can also give a noncommutative translation-invariant integral
at least on the physical subalgebra $\hat{\CC}_\lp(\R^n)$ for a finite-dimensional Lie group $G$ 
 using an operator-algebraic  completion as a Hopf $C^*$ algebra or
Hopf-von Neumann algebra. The former is based on the group
$C^*$-algebra on the group in question, which by definition is a
completion of the continuous functions on the group with
convolution product $\bullet$. This fits in with $\hat{\CC}_\lp(\R^n)$
as the image in (\ref{qgfou})-(\ref{fou}) in view of
\[ \CF(\tilde f)\CF(\tilde h)=\CF(\tilde f\bullet \tilde h),\quad \tilde f,\tilde h\in C(G).\]
  The latter Hopf-von Neumann algebra version is generated by the group-like operators (i.e. the plane waves $e_g$ as
operators on a Hilbert space)  also comes with a canonical Pentagonal operator $W=\sum e_a\tens f^a$ as an unitary
operator (see \cite{Ma:book} for an introduction) which serves as the kernel for the quantum group Fourier transform. In
either case one has a natural invariant integral in the role of Haar measure as part of the established theory.  It is
not our scope to go into details of functional analysis but we have such a completion ultimately in mind. At a more
practical level the integral in the operator algebra setting boils down to a trace in the left regular representation
after assuming a Peter-Weyl decomposition of the group coordinate algebra (and hence its dual) into matrix blocks
labelled by the irreducible representations. On any finite dimensional Hopf algebra the translation-invariant integral is
proportional to the trace composed with $S^2$ in the left regular representation \cite{Ma:book}, where $S$ is the
antipode, which means formally just the trace as $S^2$ is trivial in the classical group case. For example, if one is
thinking of the model as having momentum in $SU_2$ then up to a normalisation we have the integral on
$\hat{\CC}_\lp(\R^3)$ given by
\be\label{intSU}\int f(\hat x)=\sum_{j\in \NN}d_j
\chi_j(f(\hx))\ee
 where $d_j=j+1$ is the dimension of the representation and $\chi_j$ is the trace in the same.
 In the $SO_3$ case the integral on  $\hat{\CC}_\lp^+(\R^3)$ is similarly
\be\label{intSO} \int_+ f(\hat x)=\sum_{j\in 2\NN}d_j
\chi_j(f(\hx)).\ee Moreover, such trace formulations are
equivalent to the alternative definition \be\label{intdistrib}\int
e_{g}=\delta_1(g)\ee understood distributionally with respect to
$g$ running over the group in question (1 denotes the group
identity). Similarly for other groups. This underlies, for example, the Hopf-von Neumann
algebra treatment for any locally compact topological group.

One may expect that  $\int_+$ on $\hat{\CC}_\lp^+(\R^3)$ is in agreement
with
  $\int_\phi$ up to a normalisation
    since the coproduct essentially underlying the subalgebra is the restriction of the one essentially underlying the larger algebra. 
Thus (\ref{intdistrib}) and (\ref{intstar}) in the diagram (\ref{fou}) implies when correctly normalised
 \be\label{key} \delta_1(g)=\int_{+}
e_g={1\over 8\pi\lp^3}\int_{\R^3}\extd^3X\, \phi(e_g)=\int_{\R^3}
\frac{\extd^3X}{8\pi\lp^3} E_g \ee
 understood distributionally with respect to $g\in SO_3$, which is a key identity proven directly in \cite{PRI} and which is needed to find the right normalisation. 
 
In general the converse might not be true, i.e.  one can use  integrals defined by the same methods as in (\ref{intSU})-(\ref{intSO}) and these might extend to on a larger subspace 
$\hat{C}_\lp^I(\R^n)\subseteq \hat{C}_\lp(\R^n)$ of `integrable' functions where the sums still converge absolutely, 
but {\em a priori} we do not know they will be translation-invariant on these more general elements.  In our case since
$\hat C_\lp(\R^3)$ is not significantly bigger this is not really an issue. In fact $\int_+$ sees only the same
information as $\phi$, i.e. tied to $SO_3$ and vanishes for example on $\zeta-1$.

\subsection{Inverse quantum group Fourier transform}

Having fixed our integral $\int$ on $\hat{\CC}_\lp(\R^n)$ we can
now  write down the basic level of the quantum group Fourier
transform inverse to (\ref{qgfou}) before topological completions.
As with the forward direction (\ref{qgfou}) the construction works
similarly for any Lie group $G$ equipped with an integral $\int$
in the appropriate image $\hat{\CC}_\lp(\R^n)\subset \hat{C}_\lp(\R^n)$ and
characterised by (\ref{intdistrib}). Then from the general theory
\cite{Ma:book} adapted to this setting the inverse is
\be\label{qgfouinvsimple} 
\CF^{-1}(f)(g)=\int fe_{g^{-1}}  ,\quad
\forall  f\in \hat{C}_\lp(\R^n)
\ee 
Or in more abstract notation
\be\label{qgfouinv} \CF^{-1}(f)=\int_{G} \extd
g\,\delta_{g^{-1}}\left(\int fe_g \right)  \in C(G),\quad \forall
f\in \hat{C}_\lp(\R^3)
\ee 
where the locations $\delta_{g^{-1}}$ are
being integrated over. In the general theory on a quantum group
the kernel for the Fourier transform is the canonical element
$\sum_a e_a\tens f^a$ where $\{e_a\}$ is a linear basis of the
quantum group and $\{f_a\}$ its dual. In our case we are taking a
basis $\{\delta_g\}$ of functions on the group and dual basis
$\{e_{g}\}$ viewed as an exponential in $\hat{C}_\lp(\R^n)$, and sum
replaced by Haar integral.  On noncommutative plane waves
(\ref{qgfouinv}) becomes 
\be\label{fouinv}
\CF^{-1}(e_g)=\int_{G}\extd
h\, \delta_{h^{-1}}\left(\int e_ge_h\right)=\int_{G}\extd h\,
\delta_{h^{-1}}\delta_{e}(gh)=\delta_g\ee as required. Note that
we have used in the proof only the distributional form of the
integral (\ref{intdistrib}),  i.e. the transform and its inverse
can be verified quite directly  at this level before adding the
operator theory and functional analysis.

One can furthermore view $\CF^{-1}$ defined more generally some
suitable subspace $\hat C^I_\lp(\R^n)\subseteq \hat{C}_\lp(\R^n)$ of
`integrable' functions where   integrals of the form
(\ref{qgfouinvsimple}) converge absolutely, so that
\[ \CF^{-1}:\hat C^I_\lp(\R^n)\to C(G).\]
In this case of course it is no longer inverse $\CF$ viewed
similarly $\CF:C(G)\to \hat C^I_\lp(\R^n)$.  Rather one has
\[ \CF\circ\CF^{-1}=\CG,\quad \CF^{-1}\CF=\id\]
where 
\be\label{G} \CG: \hat C^I_\lp(\R^n)\to \hat{\CC}_\lp(\R^n),\quad
\CG(f)=\int \extd g\,\, e_{g} \left(\int f e_{g^{-1}}\right). 
\ee
  $\CG$ is a projector with image  $\hat{\CC}_\lp(\R^n)$. It corresponds to the kernel
\be \label{GG} \widetilde{\CG} \equiv \int \extd g
\,\,e_{g}\otimes e_{g^{-1}} \quad \in  \hat{\CC}_\lp(\R^n)\otimes
\hat{\CC}_\lp(\R^n). 
\ee 
Remarkably  this projection is the identity in the $SU_2$ case thanks to eq. (\ref{ek}) which states that
$$e_{[\vec{k}]}= \CG(e_{\vec{k}})= e_{\vec{k}},$$
where for $\vec k\in \R^3$ we write $\vec k\equiv k \hat k$ with $\hat k$
a unit vector in the upper `hemisphere' and $k\in \R$. We recall that $[\vec k]=\vec k -{2\pi\over \lp}n_k\hat k$ where $n_k$ is an integer and  $|[\vec k]|\le {\pi\over\lp}$.  
For $SO_{3}$  we define  $[\vec k]^+=\vec k -{\pi\over \lp}n_k\hat k$ where $n_k$
is an integer and $|[\vec k]^+|<{\pi\over 2\lp}$. 
The projection operator can be directly computed
\be
\CG^+(e_{\vec{k}})= \int_{SU_{2}}\extd g \, e_{g} \left(\int_{+}e_{g^{-1}}e_{\vec{k}}\right)
= \frac{(1+\zeta)}{2} e_{[\vec k]^+},
\ee
and just amounts to multiplication by the central projector element ${(1+\zeta)}/{2}$.

\subsection{$SO_3$ group Fourier theory}

The quantum group Fourier transform has its roots in established
Hopf algebra and operator algebra theory. If one apply the map
$\phi$ to it one obtains the composite `group Fourier
transform'  $F=\phi\circ\CF$ which
provides a very interesting and self-contained realisation with
image a subalgebra $\CC(\R^n)$ of  ordinary functions on $\R^n$ equipped with a star product.
The plane waves $e_g$ become replaced by classical fields
$\phi(e_g)$ and so forth. We have seen over the previous sections
that this indeed matches up with the quantum-gravity computations
in \cite{EL} for the group $SO_3$ (where the theory is developed
directly) for a suitable choice of $\phi$. It is also clear that
the `group Fourier theory' level provides tools of classical
analysis as well as physical insight. We illustrate these two
aspects now.

First, we describe the subalgebra  $\CC_\lp^+(\R^3)$ as the subalgebra
of the usual fields $C(\R^3)$ which is the image of $F$ in the
$SO_3$ case. As explained in \cite{EL} we can characterise it by
means of the  projection operator with kernel
 \be \label{proj} G(X,Y) =
\int_{\lp|\vec P|\leq 1} \frac{d^3\vec{P}}{(2\pi)^3}
e^{-i(\vec{X}-\vec{Y})\cdot\vec{P}} \ee satisfying \be\int
\extd^3X\, G(X,Y)G(Y,Z)=G(X,Z). \ee Then $G(f)$ is defined for all
$f\in C^I(\R^3)$ (the functions absolutely integrable on $\R^3$)
and it is shown in \cite{EL} that the image of $G$ is the same as
the image of $F$, i.e. the required space $\CC_\lp^+(\R^3)$.

The inverse group Fourier transform  can then be explicitly
written 
\be F^{-1}: C^I(\R^3) \to C(SO_3) ,\quad {f}\to
\int_{\R^3} \frac{d^3X}{8\pi \lp^3}\,\, f(X)
\der e^{\frac{1}{2\lp} \mathrm{Tr}(Xg^{-1})}
 \ee
where 
\be\label{der1}\der\equiv \sqrt{1+\lp^2 \del^i\del_i} \ee
 acts on the plane waves by  multiplication by
$P_0$ \cite{EL} (this operator can also be written $\der = 1-\imath\lp\del_0 $ if one uses  the classical
analogue of the $\hat\del_0$ operator (\ref{del0})) and 
\be
\label{invf} F^{-1}F = \mathrm{id},\quad FF^{-1}=G. \ee
 The direct proof of the first of (\ref{invf}) in \cite{EL} uses the identity (\ref{key}) just as in the quantum group case,
  but $F^{-1}$ itself does not look immediately like the realisation under $\phi$ of the quantum group  inverse transform (\ref{qgfouinvsimple}). 
  For that we need the identity
  \be\label{integralstar}
\int_{\R^3} \extd^3X (f\star g)(X) = \int_{\R^3} \extd^3X f(X)
(\der g)(X), \ee at least for all $f,g \in
\CC_\lp^+(\R^3)$.  This can be seen by the plane wave methods in
\cite{EL}. Indeed,
\begin{eqnarray*}
&&\kern-20pt \int_{\R^3}{\extd^3X\over8\pi\lp^3}E_{g_1}(X)\der E_{g_2}(X)=P_0(g_2)\int_{\R^3}{\extd^3X\over8\pi\lp^3}e^{\imath (P(g_1)+P(g_2))\cdot X}\\
&&=P_0(g_2){\pi^2\over\lp^3}\delta_0^{(3)}(\vec P(g_1)+\vec
P(g_2))=\delta_1(g_1g_2)= \int_{\RR^3} \frac{d^3X}{8\pi
\lp^3}E_{g_1g_2}\end{eqnarray*} using (\ref{key}) for the last equality and to justify
the penultimate step we need the expression \be \extd g=
\frac{l_{p}^3}{\pi^2}\frac{\extd^3P}{P_{0}}. \ee of the normalised
$SO(3)$ Haar measure  in terms of the parametrisation  (\ref{gp})
and the identity  $E_g^{-1}=E_{g^{-1}}$.

Moreover it is shown in \cite{EL} that $F$ is not only an
isomorphism but an isometry between
 $C(SO_3)\subset L^2(SO_{3})$ equipped with the normalised Haar measure and
$\CC_\lp^+(\R^3)$ equipped   with the   norm \be ||f||_{\lp}^2 \equiv
 \int_{\RR^3} \frac{d^3X}{8\pi \lp^3} (f\star \bar{f})(X).
\ee Clearly $G\circ\phi=\phi\circ\CG$ to the extent that $\phi$ is
defined and we have the same remarks as for $\CG$ mentioned above. 

\subsection{$4$-dimensional $SU_2$ Fourier transform}

 We recall that the noncommutative calculus on $\hat{C}_\lp(\R^3)$ has a fourth direction $\hd_0$ with conjugate 1-form
$\theta$ given by the identify matrix in the Pauli algebra, this needs to be added to the other Pauli matrices. We have
seen that the operator plays a mild if not very explicit role in the $SO_3$ version of 2+1 quantum gravity but not the
need for an extra coordinate to match this extra `direction'. We now look at the $SU_2$ version of the theory which
until now has not been fit into the group Fourier transform framework. We find that it can be done precisely if one
indeed adjoins a new variable $T$ as suggested by the noncommutative geometry.  We shall argue that its meaning for 2+1
quantum gravity is not `time' cf \cite{Ma:time} but rather that hints at a unification of geometry with the
renormalisation group.

 First, if one tries to prove the key identity (\ref{key}) needed to try to have a  `group Fourier theory' on $SU_2$ using functions $C(\R^3)$, one has \cite{PRI}  the $ SU_{2}$ $\delta$-distribution  at the group identity expressed as an integral
\be\label{deltaid2} \delta_1(g)=\int_{\R^3} \frac{\extd^3X}{4\pi
l_{p^3}}e^{\frac{1}{2\lp}\mathrm{Tr}(Xg)} \Theta(\mathrm{Tr}(g))
\ee
 where $\Theta(X)$ is the Heaviside distribution $\Theta(x)=0, \, x<0, \, \Theta(x)=1, \, x>0$. But one cannot simply consider  $E_g\Theta({\rm Tr}(g))$ as a `new plane wave' suitable for $SU_2$ since it is not invertible and does not lead to any meaningful $\star$-product.

 We can however, express the Heaviside distribution as an integral
 \be
 \Theta(X)= \frac{1}{2i\pi} \int_{\RR} \frac{\extd T}{T-i0} e^{iT X}
 \ee
 where we mean to take a contour of
 integration  along the real line which goes below  the singularity at $T=0$.
 We can now write the $SU_{2}$ delta function in a form suitable for us, in terms of a $4$-dimensional integral.
 We introduce the quadrivectors $P_{\mu}=(P_{0}, \vec{P})$, $X^{\mu}=(X^{0}, \vec{X})$ and the measure
 \be
 \widetilde{\extd^4X}\equiv \frac{\extd^3X}{4\pi {\lp}^3}\frac{\extd T}{2i\pi (T-i0)}
 \ee
where the $T=X^0$ integration should always be done before the
$\vec{X}$, then
 \be \label{key-}
 \delta_1(g)=\int_{\RR^4}\widetilde{\extd^4X}
 e^{\imath X^\mu P_{\mu}} = \int e_{g}\ee
 for the $SU_2$ model (\ref{intSU}).

 The plane waves are now $4$-dimensional
 \be
 \wE_{g}(X)=  e^{\imath X^\mu P_{\mu}(g)}=e^{\frac{1}{2\lp}\mathrm{Tr}(Xg)}
 \ee
 where now $X=X_0{\rm id}+X^i\sigma_i$ is the extension of the Pauli matrix
 representation in parallel with the extension of the calculus.
Since we have extended the spacetime to $4$ dimensions it is also
natural to extend the group structure. The natural extension is to
consider a trivial central extension of $SU_2$ denoted
$\widetilde{SU}_2 = \RR^+ \times SU_2$ which is the space of two
by two matrices of the form $ \tilde{g}\equiv \alpha g$ where
$\alpha >0$ and $g \in SU_2$. In other words \be \tilde{g}=
P_0 \id + \imath \lp P^i\sigma_i,\quad P_0^2 +\lp^2 P^iP_i>0.
\ee
We  now  define a $\star$ product  by the group product  as before, which is to say
\be
\wE_{\alpha_{1} g_{2}} \tilde{\star} \wE_{\alpha_{2} g_{2}}= \wE_{\alpha_{1}\alpha_{2}g_{1}g_{2}}
\ee
or
 \be
  e^{\imath X^\mu P_{\mu}}\tilde{\star}  e^{\imath X^\mu Q_{\mu}}
  =  e^{\imath X^\mu (P\oplus Q)_{\mu}},
  \ee
  with
  \be
  (P\oplus Q)_{0} = P_{0}Q_{0} - \lp^2\vec{P}\cdot\vec{Q}, \quad
  (P\oplus Q)_{i} = P_{0}Q_{i} +Q_{0}P_{i}- l_{P} \vec{P}\times \vec{Q}.
 \ee
 Note that the identity for this product is not given by the identity function but by $\exp(iT)$, 
 the action of functions of $T$ is given by
 \be\label{Tprod}
 e^{i\alpha T}\tilde{\star}  e^{i\beta T} = e^{i\alpha \beta T},\quad 
 e^{i\alpha T}\tilde{\star}  \wE_{g}(X)= \wE_{\alpha g}(X).
 \ee
This product  yields a 4-dimensional deformation of $C(\R^4)$  provided we expand functions around 
the star product identity $\exp(\imath T)$. Thus 
\[ e^{\imath T}T\tilde\star e^{\imath T}T=e^{\imath T}(T^2-\imath T),\quad e^{\imath T}X^i\tilde\star e^{\imath T}X^j= e^{\imath T}(X_iX_j+\imath\lp\eps^{ij}{}_k X_k+\lp^2\delta_{ij}T)\]
\be\label{4drelns} e^{\imath T}T\tilde\star e^{\imath T}X_i= e^{\imath T}X^i\tilde\star e^{\imath T}T=e^{\imath T}(TX_i-\imath X_i)\ee
so that elements of the form $e^{\imath T}C^{\rm poly}(\R^4)$ induce a $\star$-product deformation of $C^{\rm poly}(\R^4)$.
We assume that the $\tilde\star$ algebra extends to general functions and denote it by $\hat C_\lp(\R^4)$; in view of the above it is a central extension of our previous
$\hat C_\lp(\R^3)$ by an additional central `time' generator when expanded correctly.

The group Fourier transform is
defined to be 
\be\label{Fou4}\displaystyle
F(\tilde{f})(X) = \int_{SU_{2}} \extd g  \wE_{g}(X) \tilde{f}(g),
\ee
 where $F:  C(SU_{2})\to\CC_\lp(\RR^4)$ and $\CC_\lp(\RR^4)$ is defined
to be the image of  $F$. One can check that $f\in C(\RR^4)$  lies
in $\CC_\lp(\R^4)$  if and only if $f|_{T=0}\in \CC_\lp(\RR^3)$ and the
time dependence is controlled by a Laplace equation 
\be
\label{Lapl} (\partial_{T}{\partial_T}+
\lp^2\partial_{i}\partial^{i})f= -f. 
\ee

Conversely if we consider a function $f$ in $\CC_\lp(\RR^3)$ or more generally in $C(\RR^3)$
we can construct two functions on $\CC_\lp(\RR^4)$ by convolution with $G_{\pm}$
\bes
G_{\pm}(f)(X)&\equiv&\int \extd^3Y G_{\pm}({X},\vec{Y})f(\vec{Y}), \\ 
G_{{\pm}}(X,\vec{Y})&=&
\int_{\lp|\vec{P}|<1}\frac{\extd^3P}{(2\pi)^{3}}e^{\pm \imath
T\sqrt{1-|\vec P|^2} }e^{- \imath \vec{P}\cdot(\vec{X}-\vec{Y})} 
\ees
which project on the positive and negative 'energy' solutions of
(\ref{Lapl}) and $X$ denote the $4$-vector $(T,\vec{X})$.
This gives an isomorphism 
\bes \label{posneg}
 \CC_\lp(\RR^3)\oplus\CC_\lp(\RR^3) &\rightarrow& \CC_\lp(\RR^4) \\
 (f_{+},f_{-}) &\mapsto & G_{+}(f_{+}) +G_{-}(f_{-}).\nonumber
 \ees
  In order to show that this map is invertible lets consider $F\in \CC_\lp(\RR^4)$, If $F = G_{+}(f_{+}) +G_{-}(f_{-})$
  with $f,g \in \CC_\lp(\RR^3)$ then 
  \be 
 [ (-\imath \partial_{T}+ \der)F](0,\vec{X})= 2\der f_{+}(\vec{X}),\quad  [ (\imath \partial_{T}+ \der)F](0,\vec{X})=
2\der f_{-}(\vec{X}),
 \ee
where $\der$ is defined in (\ref{der1}) and since $G_{\pm}$ are identity operators on $\CC_\lp(\RR^3)$ when restricted to the slice $T=0$.
The hermitian positive operator $\der$ is invertible on $\CC_\lp(\RR^3)$ so for a general $F$ we can define 
$2 f_{\pm} = \der^{-1} [ (\mp \imath \partial_{T}+ \der)F]_{T=0}$.
Then $\tilde{F}= F -G_{+}(f)-G_{-}(g)$ is a solution of (\ref{Lapl}) satisfying $\tilde{F}_{T=0}=0=(\partial_{T}\tilde{F})_{T=0}$
it is therefore a null function.

This shows that as a vector space
\be\label{CCR4}\CC_\lp(\RR^4)\cong \CC_\lp(\RR^3)\oplus\CC_\lp(\RR^3),\ee
the direct sum of two copies of $\CC_\lp(\RR^3)$
where $f= f_+ + f_-\in \CC_\lp(\RR^4)$ is decomposed as a sum of positive and negative energy solutions of 
(\ref{Lapl}).
We recover the $SO_{3}$ Fourier transform and $\star$ product if one restricts to 'even' functions
of $\CC_\lp(\R^4)$ which are in the image of $G= (G_{+}+G_{-})/2$.
This mapping is such that it intertwines the 4 dimensional $\star$ product with the 
three dimensional one 
\be
G(f)\tilde{\star}  G(g)= G(f\star g).
\ee
The inverse Fourier transform is given by 
\be\label{invFou3}
 \displaystyle F^{-1}({f})(g) = \int_{\R^4}  \widetilde{\extd^4X} (E_{g^{-1}} \tilde{\star} {f})(X).
\ee
The Fourier transform is an isometry between $C(SU_{2})$ equipped with the usual $L^{2}$ norm and $\CC_\lp(\RR^4)$
equipped with the norm
\be\label{norm1}
||f||^2= \int\widetilde{\extd^4 X} \bar{f}\tilde{\star}f
\ee
where $\bar{f}\equiv f^*(-T,X)$ is the combination of complex conjugation and time reversal, it is such that 
$\bar{\wE}_{g}= \wE_{g^{-1}}$. 
The fact that this scalar product is isometric to the $L^2$ norm on $SU_{2}$ can be verified directly using (\ref{key-}).

We can express this norm  in terms of the previous identification as follows
\be\label{norm2}
||G_{+}(f_{+})+G_{-}(f_{-})||^2= \int_{T=0}\frac{\extd^3 X}{4\pi \lp^3} (\bar{f}_{+}\der f_{+} +  \bar{f}_{-}\der f_{-}),
\ee
where $\bar{f}$ denotes the complex conjugation.
This shows that the decomposition (\ref{CCR4}) is in term of orthogonal subspaces.
We could define similar norms $||\cdot||_{a}$ using a time slice $T=a$ instead of $T=0$, this however doesn't lead to a new norm since 
 $||\cdot||_{a}= ||\cdot ||$ for any $a$.
 
 In order to prove these statements we need to establish few lemmas.
 First,  lets consider $f\in \CC_\lp(\RR^4)$ and lets denote $f_{\pm}\equiv G_{\pm}(g_{\pm})$ its positive and negative energy 
 components. $f$ is in the image under the group Fourier transform (\ref{Fou4}) of $\tilde{f}(g)$. The identification of
$SU_{2}$
 with the $3$-sphere $P_{0}^2 +\lp^2\vec{P}^2=1 $ means that function on $SU_{2}$
 is determined by $\widetilde{f}_{\pm}(\vec{P})\equiv \tilde{f}(\pm\sqrt{1-\lp^2 \vec{P}^2}, \vec{P})$.
 We can express the $L^2$ norm on $SU_{2}$ in terms of these variables
 \be
 \int_{SU_{2}}\extd g \bar{\tilde{f}}(g)\tilde{f}(g)=\int_{\lp |\vec{P}|<1} \frac{\lp^3 \extd^3P}{2\pi^2\sqrt{1- \lp^{2}\vec{P}^2}}
 (\bar{\tilde{f}}_{+}{\tilde{f}}_{+}+ \bar{\tilde{f}}_{-}{\tilde{f}}_{-})(\vec{P}).
 \ee
 The Fourier transformation (\ref{Fou4}) can also be written in this variables as 
 \be\label{Fou5}
f_{\pm}(X)=\int_{\lp |\vec{P}|<1} \frac{\lp^3 \extd^3P}{2\pi^2}
\frac{e^{\pm \imath P_{0}T}}{P_{0}} e^{\imath\vec{P}\cdot\vec{X}} \tilde{f}_{\pm}(\vec{P})
\ee
where  $P_{0}=\sqrt{1- \lp^{2}\vec{P}^2}$.
This can be easily inverted by the usual inverse Fourier transform as 
\be\label{invFou4}
\tilde{f}_{\pm}(\vec{P})= 
\int_{T} \frac{\extd X^3}{4\pi \lp^3}e^{\mp \imath P_{0}T} e^{-\imath \vec{P}\cdot\vec{X}} (\der f)(X)
 \ee
 where the integral is over a three dimensional slice $T=\mathrm{constant}$ (but not necessarily $T=0$).
 This gives an alternative but equivalent formula for the inverse Fourier transform (\ref{invFou3}).
 It is now a straightforward computation to show that 
 \be
 \int_{\lp |\vec{P}|<1} \frac{\lp^3 \extd^3P}{2\pi^2\sqrt{1- \lp^{2}\vec{P}^2} } \bar{\tilde{f}}_{\pm}{\tilde{f}}_{\pm}(\vec{P})
 = \int_{T} \frac{\extd X^3}{4\pi \lp^3} (\bar{f}_{\pm}\der f_{\pm})(X),
 \ee
 which proves our claims.

We can now give a full $4$ dimensional perspective to our construction if one introduce 
the Green function $P=(P_{+}+P_{-})/2$ where 
\be
P_{{\pm}}(X^{\mu})=
\int_{\lp|\vec{P}|<1}\frac{\lp^3 \extd^3P}{2\pi^{2}\sqrt{1-\lp^2\vec{P}^2}}e^{\pm \imath
T\sqrt{1-\lp^2\vec{P}^2} }e^{- \imath \vec{P}\cdot\vec{X}}.
\ee
 Note the key factor $\sqrt{1-\lp^2\vec{P}^2}$ in the integrand compared to the definition of $G$.
 We consider the following sesquilinear form  on  $C(\RR^4)$ the space of all functions on $\RR^4$ (more precisely 
 a dense subspace of $L^2$ functions on $\RR^4$ like Schwartz space),
given by 
\be \label{norm3}
\langle f|f\rangle = \int\extd^4X \extd^4Y\, \bar{f}(X)P(X-Y){f}(Y).
\ee
We will show that this bilinear form is positive, however it is not definite; it possess a kernel
$\mathrm{Ker}\equiv \{f\in C(\RR^4)| \,\langle f|g\rangle=0\,\, \forall g \in  C(\RR^4)\}$.
Since the form is positive we have
 $\mathrm{Ker}= \{f\in C(\RR^4)| \,\langle f|f\rangle=0 \}$.
 On $C(\RR^4)$ we impose the equivalence relation $f\sim 0$ if $f\in \mathrm{Ker}$
 and we consider the quotient space $C(\RR^4) / \sim$.
 The GNS construction ensure that this space is  a Hilbert space with the induced norm between equivalence class
 $\langle [f]|[f]\rangle\equiv\langle f|f\rangle$. 
 Our main claim is now that  $C(\RR^4) / \sim$ is isomorphic as an Hilbert space with $\CC_\lp(\RR^4)$ equipped with the norm 
 (\ref{norm1},\ref{norm2}).
 
 In order to see that the form (\ref{norm3}) is positive it is convenient to write it in terms of the $4$-dimensional Fourier modes
 \be
 \tilde{f}(P)= \int \extd^{4}X e^{iP\cdot X} f(X).
 \ee
One easily sees that 
\be \label{norm4}
\langle f|f\rangle =\int \frac{\lp^3\extd^4P}{\pi^2}\delta(P_{0}^2+\lp^2\vec{P}^2-1) \bar{\tilde{f}} (P)\tilde{f}(P)
\ee
where we recognise on the RHS the normalised integral on the momentum space $3$-sphere $P_{0}^2+\lp^2\vec{P}^2=1$ which is positive. From this expression it is clear that $\mathrm{Ker}$ is generated by all functions such that $\tilde{f}|_{P_{0}^2+\lp^2\vec{P}^2=1}=0$.
Equivalently we can show that  $\mathrm{Ker} = \mathrm{Im}(\partial_{T}^2+ \der^2)$. Therefore the quotient space 
$C(\RR^4)/\mathrm{Ker} = \mathrm{Ker}(\partial_{T}^2+ \der^2)= \CC_\lp(\RR^4)$. Moreover the norm on this quotient 
space is given by (\ref{norm4}) which is the $L^2$ norm on $SU_{2}$, this finishes the proof.

\section{Radial functions, Gaussian functions and Duflo map}

Now that we have a developed general tools of `noncommutative' harmonic analysis for $SO_3$ and $SU_2$, we now  study in more detail the space of radial functions which are invariant under the rotation group and where we can take calculations much further.

\subsection{Orbit integral }

As we have already seen in section (\ref{starprod}) $SU_{2}$ acts naturally by rotation on $\hat{C}_{\lp}(\RR^3)$,
moreover if one average over all $SU_{2}$ this action we obtain a map  
\bes \label{Rad}
 \hat{C}_{\lp}(\RR^3)&\rightarrow& Z_{\lp}(\RR^3) \\
 f &\mapsto & R(f)\equiv \int_{SU_{2}} \extd g \, g\la f,\nonumber
 \ees
where $Z_{\lp}(\RR^3) $ denotes the centre of $ \hat{C}_{\lp}(\RR^3)$.
Note that the action by rotation can be obtained by the adjoint action of plane wave 
\be
g\la f= e_{g} f e_{g}^{-1}.
\ee
The image of $R$ is a 'radial' function which depends on $\hx^{i}$ only  through the quadratic casimir $\hat{c}= \hx^{i}\hx_{i}$
and it is obviously a central function. Of course, any radial function is in the image of $R$ since $R$ is the identity on them.
If $f\in \hat{\CC}_{\lp}(\RR^3)$ the quantum group fourier components of $R(f)$ are class functions on $SU_{2}$ and $R(f)$ depends on the plane wave 
through $R(e_{g})$. In order to study the properties of the space of radial functions and its image under $\phi$ it is therefore natural 
to look more closely at the properties of $R(e_{g})$ and $\phi(R(e_{g}))$.

In order to do so lets introduce some notations: Given $e_{g}=e^{i {k}_{i}\cdot \hx^{i}}$ we  denote 
\[\theta \equiv l_{p} |\vec k|\]
we also introduce   radial variables
\be
 \hr^2\equiv {\hx_{i}\hx^{i}}+ {\lp^2},\quad r^2\equiv X_{i}X^{i},
 \ee
 and dimensionless radial variables 
 \be
 \hrho \equiv \frac{\hr}{\lp},\quad \rho\equiv \frac{r}{\lp}.
 \ee
We shall prove now the following:
\bes \label{orbit1}
R(e_{g}) &=& \int_{SU_{2}}\extd h \, e_{hgh^{-1}} = \frac{\sin (\hrho \theta)}{\hrho \sin \theta}, \\\label{orbit2}
\phi\left(R(e_{g})\right) &=& \int_{SU_{2}}\extd h \,E_{hgh^{-1}} = \frac{\sin( \rho\sin \theta)}{\rho\sin \theta}.
\ees
The first equality is valid as long as $\theta<\pi$ but by continuity extends to $\theta=\pi$. 
An immediate corollary is that  
\be
\label{sinr}\sin\left({\pi\hr\over\lp}\right)=0, \quad \zeta = -\cos\left({\pi\hr\over\lp}\right)
\ee
 for consistency of the limit.  Another immediate corollary of (\ref{orbit2}) is the identity valid when  $\theta\le \pi$
\be
\label{rphi} \phi\left(\frac{\sin(\hrho  \theta )}{\hrho}\right)=\frac{\sin( \rho{\sin\theta})}{\rho}.
\ee
{} From this it follows that
 \be\label{phir}\phi\left(\hr^{2m}\right)=\sum_{n=0}^{m}\lp^{2(m-n)} C_{n,2m+1}r^{2n},\ee
 with $C_{n,m}$ defined in  section \ref{starprod}. 
  
 The proof of (\ref{orbit2}) is done by a direct computation of the integral. In order to prove (\ref{orbit1}) lets compute the trace 
 of the LHS in the representation of weight $j$, since $\chi_{j}$ is a class function we have 
 \be
 \chi_{j}(R(e_{g}))= \chi_{j}(e_{g}) = \chi_{j}(g)= \frac{\sin(j+1) \theta}{\sin{\theta}}.
 \ee 
 Any even function of $\hr$ function acts diagonally  on the representation of weight $j$, $ f(\hr) V_{j}= f(\lp(j+1)) V_{j}$.
 So the trace RHS of (\ref{orbit1}) in the representation of weight $j$ is given by
 \be
  \frac{\sin ((j+1)\theta)}{(j+1)\sin \theta} \chi_{j}(1)= \frac{\sin ((j+1)\theta)}{\sin \theta}.
  \ee
{}From this we can conclude that the LHS and RHS of (\ref{orbit1}) agree when evaluated in any finite dimension representation of (\ref{su2}).
We can expand both sides of (\ref{orbit1}) as a series in $\theta$, at each order in the expansion the coefficients of $\theta^{n}$ is 
a polynomial in $x^{i}$ of degree at most $n$ invariant under the action  of $SU_{2}$. The LHS and RHS polynomials take the same value on 
integers by the previous reasoning, they are thus identical. This proves that the Taylor series in $\theta$ agree. Now one can easily see that the Taylor 
expansion of the RHS of (\ref{orbit1}) has a radius of convergence $\pi$. This shows that (\ref{orbit1}) is valid as long as 
$\lp |\vec k|<\pi$ as claimed.
In order to get (\ref{sinr}) we use the fact that the limit $e_{g}\to e_{{-1}}= \zeta$ or equivalently $\theta \to \pi$
is
well defined, since $\zeta$ is central, this implies that the RHS is just $\zeta$. Identities (\ref{sinr}) are obtained by
taking the limit  $\theta \to \pi$ in the RHS.

Finally, as an application, let us define `radial waves' depending
only on a modulus of the momentum $k\in[0,\frac{\pi}{\lp}[$,
\be\label{sphwave} \psi_k(\hr)=\frac{\lp\sin(\hr k)}{\hat r\sin\lp k}.\ee
Since these are given by the orbit integral of $e_{g}$, they diagonalise 
 the Laplace operator $\hd_0$ (see (\ref{del0})), with
  \be\label{rdel0}\hd_0\psi_k(\hr)={\imath\over\lp}(\cos(\lp k)-1)\psi_k(\hr),\quad \Del\psi_k(\hr)=\cos(\lp k)\psi_k(\hr).\ee 
 Note that the second form says that
  \[ \Del\left(\frac{\sin(\hr k)}{\hat r}\right)=\frac{\sin(\hr k)\cos(\lp k)}{\hat r}={\sin((\hr+\lp)k)- \sin((\hr-\lp)k)\over 2\hr }\]
  which suggests that 
  \be\label{Delr}
  \Del f(\hr) = \frac{1}{2}\left((1+\frac{\lp}{\hr})f(\hat r+\lp)+(1-\frac{\lp}{\hr})f(\hat r-\lp)\right).
\ee
 on any reasonable $f(\hr)$. This will be proven directly in Section~\ref{rcovcal}.
 
\subsection{Character expansion and radial sampling theorem}
The usual Fourier theory of $SU_{2}$  expresses a function on the group as an expansion 
in terms of characters
\be
\tilde{f}(g) =\sum_{j\in \NN} d_{j} (\chi_{j}\bullet \tilde{f}) (g)
\ee
where $d_{j}=j+1$, $\chi_{j}(g)$ is the character of the weight $j$ representation and $\bullet$ is the convolution product.
Under the quantum group fourier transform this expansion becomes 
\be\label{fhx}
{f}(\hx) =\sum_{j\in \NN} d_{j}\, \hat\delta_{j} \cdot {f},\quad\forall f\in\CC_\lp(\R^3)
\ee
where we have defined  `quantum delta-functions' $\hat\delta_{j}(\hr)$ which are radial 
 functions obtained by Fourier transform of the characters
 \[ \hat\delta_{j}(\hr)=\CF(\chi_j)=\int\extd g\, \chi_j(g) e_g\in\hat{\CC}_\lp(\R^3).\]
 Starting from (\ref{orbit1}) we can compute them explicitly
 \be
 \hat\delta_{j}=(-1)^{j+1}\left( \frac{\sin \pi x}{\pi x} \frac{2 (j+1)}{(j+1)^2-x^2}\right)|_{x=\hrho},
 \ee
 with $\hrho=\hr/\lp$.
 It follows by convolution of characters that $d_{i}\hat\delta_i$ are orthogonal projectors
\be
\label{prodel}
 \hat\delta_i\hat\delta_j={\delta_{ij}\over d_i}\hat\delta_i.
 \ee

 Geometrically these project  on the quantum sphere of radius $\hr =\lp(j+1)$. Indeed, lets consider 
$f\equiv f(\hr)$ a radial function in $\hat{\CC}_{\lp}(R^3)$,  the following key property is satisfied
\be\label{deltaj}
\hat\delta_{j} \cdot {f} = f(\lp(j+1)) \hat\delta_{j}.
\ee
Thus we have the `sampling theorem' that a radial function in $\hat\CC_\lp(\R^3)$ can be recovered from its values on integers,
\be\label{sampleth}
{f}(\hr) =\sum_{j\in \NN} d_{j}  {f}(\lp(j+1))\, \hat\delta_{j}
\ee
from (\ref{fhx}) or equivalently
\[ 1=\sum_{j\in \NN}d_j \hat\delta_j(\hr).\]

To prove these results,  note that under the quantum group inverse Fourier transform the LHS of (\ref{deltaj}) becomes 
\bes 
&\chi_{j}\bullet \tilde{f}(g) = \int_{SU_{2}} \extd h \, \chi_{j}(gh^{-1}) \tilde{f}(h)=  
\int_{SU_{2}} \extd h \extd u \, \chi_{j}(guh^{-1}u^{-1}) \tilde{f}(h)\nonumber\\
& =
\frac{\chi_{j}(g)}{d_{j}} \int_{SU_{2}} \extd h\, \chi_{j}(h) \tilde{f}(h)
= \frac{\chi_{j}(g)}{d_{j}} \chi_{j}(f(\hr)) = \chi_{j}(g) f(\lp(j+1)).\nonumber
\ees
The first equality is the definition of the convolution product, the second equality follows after a change of variables from the 
fact that $\tilde{f}$ is a class function, the third equality is obtained after integration over the $u$ variable and from the fact that $\chi_{j}(h)=\chi_{j}(h^{-1})$, the fourth equality follows from the definition 
of the quantum group Fourier transform and in the  last equality 
we make use of the fact that $f(\hr)$ acts by scalar multiplication on the representation of weight $j$.
In the last term we recognise the inverse Fourier transform of $ f(\lp(j+1)) \hat\delta_{j}$ this proves (\ref{deltaj}).

A first application of the character expansion or sampling theorem is the observation 
\be\label{zetar}
\zeta=\sum_{j\in \NN}d_j(-1)^j\hat\delta_j(\hr)
\ee
This follows from the sampling theorem since $\zeta=-\cos(\pi \hrho)$ (\ref{sinr}) and $(-1)^j= -\cos((j+1)\pi)$.
Also the Fourier transform of $\zeta=e_{-1}$ is given by the character expansion $\delta_{-1}(g)=\sum_{j\in
\NN}d_j(-1)^j \chi_{j}(g)$. One can readily check from this expression and the product (\ref{prodel}) that $\zeta^2=1$.

An additional interesting property of the `quantum delta-functions'  are their properties under integration (\ref{intSU}) on $\hat{\CC}_{\lp}(\RR^3)$ 
\be 
\int \hat\delta_{j} \cdot {f} = d_{j} R(f)(\lp(j+1)),
\ee
where $f\in\hat{\CC}_{\lp}(\RR^3) $ and $R(f)$ is the corresponding radial function (\ref{Rad}).
Since $\hat\delta_{j} $ is a radial function then the LHS is equal to $\int \hat\delta_{j} \cdot R({f})$ which can be evaluated thanks to the 
the previous property. We are left with $\int \hat\delta_{j}$ which is equal to $d_{j}$ since
\be
\chi_{i}(\hat\delta_{j})  = \int_{SU_{2}} \extd g\, \chi_{i}(g)\chi_{j}(g) = \delta_{ij}.
\ee 

Under the map $\phi$ the quantum delta-functions are mapped to the $SO_{3}$ group Fourier transform of the
characters, from a direct computation we get  \cite{EL} that
\be\label{besselfou} 
\delta_{j}(r)\equiv \phi(\hat\delta_{j} )(r)=  \int
\extd g\, \chi_j(g)E_g(X) = \left\{
\begin{array}{c}
0, \mathrm{if }\, j\, \mathrm{is\, odd}\\
\displaystyle
2\frac{ J_{j+1}(\rho)}{\rho},\, \mathrm{if}\, j\, \mathrm{is\, even}
\end{array}
\right.
\ee 
where $\rho=r/\lp$ and $J_{n}$ are  the Bessel
functions. They are defined as the  Fourier modes of
 $e^{\imath \rho\sin\theta}=\sum_{n}e^{\imath n\theta}J_{n}(\rho)$. One has
 from their definition and the properties $J_{-n}(\rho)=J_{n}(-\rho)=(-1)^n J_{n}(\rho)$,  the following identity
 \be
 \frac{\sin(\rho \sin\theta)}{\rho\sin\theta}=2 \sum_{j\in 2\NN}
 \frac{\sin(j+1)\theta}{\sin\theta}\frac{J_{j+1}(\rho)}{\rho},
 \ee
 from which we get
 \be\label{besselpart}
 1=\sum_{j\in 2\NN}
 (j+1)2\frac{J_{j+1}(\rho)}{\rho}.
 \ee
 This is consistent with our above results, for we deduce from (\ref{zetar}) that 
\[ \frac{1+\zeta}{2}=\sum_{j\in 2\NN}(j+1)\hat\delta_{j}\in \hat{\CC}_{\lp}^+(\R^3).\]
and $(1+\zeta)/2$  is indeed mapped to the identity by the map $\phi$. It is a projector from  $\hat{\CC}_{\lp}(\R^3)$ to  $\hat{\CC}_{\lp}^+(\R^3)$ and 
acts as the identity element for  the algebra $\hat{\CC}_{\lp}^+(\R^3)$.
 We also conclude since the $\hat\delta_j(\hr)$ are  a basis
for the radial functions that $\hat C_\lp(\R^3)$ splits as a vector space into $\CC^+_{\lp}(\R^3)\oplus \hat\CC_\lp^-(\R^3)$ according to this projection.

In order to have access to the odd spin delta functions we need to consider the $SU_{2}$ Fourier theory.
A direct computation  shows that 
\be\label{besselfou2} 
\tilde \delta_{j}(r,T)\equiv \int_{SU_{2}}\extd g\, \chi_j(g)\widetilde{E}_g(X) = \left\{
\begin{array}{c}
\displaystyle 2\imath\frac{ J_{j+1}(\sqrt{\rho^2 +T^2})}{\rho} \sin(j+1)\varphi, \mathrm{if }\, j\, \mathrm{is\, odd}\\
  {}\\
\displaystyle
2\frac{ J_{j+1}(\sqrt{\rho^2 +T^2})}{\rho} \cos(j+1)\varphi,\, \mathrm{if}\, j\, \mathrm{is\, even}
\end{array}
\right.
\ee 
with $ e^{\imath\varphi}\equiv \frac{\rho+iT}{\sqrt{\rho^2 +T^2}}$. From this definition we can easily check that 
\be
\cos T =\sum_{j \in 2\NN} (j+1) \tilde \delta_{j}(r,T),\quad \imath \sin T =\sum_{j \in 2\NN +1} (j+1) \tilde \delta_{j}(r,T),
\ee
these are the projector onto the odd and even subspace of $\CC_\lp(\RR^4)$ since 
\be 
\cos T\tilde \star \widetilde{E}_{g}= \frac{1}{2}(\widetilde{E}_{g}+\widetilde{E}_{-g}), \quad
\sin T\tilde \star \widetilde{E}_{g}= \frac{\imath}{2}(\widetilde{E}_{g}-\widetilde{E}_{-g}).
\ee
More generally  we have 
\bes \displaystyle
\cos(T \cos \theta) \frac{\sin(\rho\sin \theta)}{\rho\sin \theta}&=&\sum_{j \in 2\NN} \tilde \delta_{j}(r,T) \frac{\sin(j+1)\theta}{\sin \theta},\\
\displaystyle
\quad \imath\sin(T \cos \theta) \frac{\sin(\rho\sin \theta)}{\rho\sin \theta}&=&\sum_{j \in 2\NN +1} \tilde \delta_{j}(r,T) \frac{\sin(j+1)\theta}{\sin \theta}.
\ees
These formulae make it clear that $e^{-\imath T}$ plays the role of $\zeta\in\hat\CC_\lp(\R^3)$ (in other words the classicalisation of this now possible using our $SU_2$ group Fourier theory) while as mentioned already the classicalisation of $1$ in $\hat\CC_\lp(\R^3)$  is $e^{\imath T}$. The star product relation $e^{-\imath T}\tilde\star e^{-\imath T}$ plays the role of $\zeta^2=1$.

\subsection{Duflo map}
Given a function $f\in C(\RR^3)$ we define the `extended Duflo map' to be the quantisation map ${\DD}:C(\RR^3) \to \hat{C}_{\lp}(\RR^3) $ defined by (\ref{Duflowave}) 
on the plane waves. This extends linearly to any function
\be\label{fk}
f(\vec{X}) = \int_{\RR^3} \frac{\extd^{3}k}{(2\pi)^3}e^{\imath\vec{k}\cdot\vec{X}} \tilde f (\vec{k}).
\ee
as 
\be\label{duflo}
\DD(f) = \int_{\RR^3} \frac{\extd^{3}k}{(2\pi)^3}\frac{\sin\lp |\vec{k}|}{\lp |\vec{k}|}
e^{\imath k_{i}\hx^{i}}\tilde f (\vec{k})
\ee
and maps over under the two Fourier isomorphisms in (\ref{SU2diag}) to
\be\label{mapp} p(\tilde f)(\vec k)= \sum_{n\in \ZZ} \frac{\lp k+2n {\pi}}{\sin (\lp k)} \tilde{f}\left(
(k+2n\frac{\pi}{\lp})\hat{k}\right),\quad |k|\le {\pi\over\lp}\ee
with $\vec k\equiv k \hat k$, where $\hat k$ is a unit vector in the upper `hemisphere' and $k\in \R$. One has 
\be\label{compress1}\CF(p(\tilde f))=\int_{|\vec k|\le{\pi\over\lp}}\extd^3k \left(\frac{\sin(\lp |\vec k|)}{\lp |\vec k|}\right)^2 
e^{\imath\vec{k}\cdot\hx}p (\tilde f)(\vec k)
=\DD(f)\ee
as required. The proof follows from  the identity (\ref{ek}): The integral over $\vec{k}$ 
expressing $\DD(f)$ can be expanded as a sum over $n$ of integrals over $k\in  ]-{\pi\over \lp}, {\pi\over\lp}[+{2n\pi\over\lp}$  (with a convention on the boundary as in Section~1) and after 
a change of variables, including the change in measure, can be expressed as the quantum group Fourier transform stated. Next, we make $p$ into a projection by composing back with a map   
\[ {\rm extn}(\tilde f)(\vec k)=\begin{cases}\tilde f(\vec k){\sin \lp |\vec k|\over \lp |\vec k|} & {\rm if}\ |\vec k|<{\pi\over\lp} \\ 0& {\rm else}\end{cases}\]
which extends by zero, but with a suitable weight so that ${\rm extn.}$ followed by $p$ is the identity. The  resulting  projection operator $ \tilde\Pi={\rm extn.}\circ p$  on $C'(\R^3)$ is 
\be\label{compress2}
\tilde{\Pi}(\tilde f )(\vec{k}) =\sum_{n\in \ZZ}\frac{\lp k +2n \pi}{\lp k}
\tilde{f}\left((k+2n\frac{\pi}{\lp})\hat{k}\right).
\ee
Under the Fourier isomorphisms in (\ref{SU2diag}) the map ${\rm extn}$ maps over to a map $i$ which on noncommutative plane waves is
\be\label{mapi} i(e^{\imath \vec k\cdot\hx})=e^{\imath \vec k\cdot X}{\lp |\vec k|\over\sin \lp |\vec k|}\ee 
and clearly $i$ followed by $\DD$ is the identity. The resulting projector $\Pi=i\circ\DD$ on $C(\R^3)$ comes out  as  
\be\label{Piwave} \Pi(e^{\imath \vec k\cdot X})={|[\vec k]|\over |\vec k|}e^{\imath [\vec k]\cdot X}\ee
in the same notation as in (\ref{ek}), modulo technical choices when $|\vec k|=\pi/\lp$ since the map $i$ is singular there. Compare with the circle case in Section~1. This completes the proof of the commutative diagram (\ref{SU2diag}) as determined once the Duflo map $\CD$ is fixed as in (\ref{duflo}). 
By construction, the two `noncommutative compression maps' $\tilde \Pi,\Pi$ are projections  related by the $\R^3$ Fourier transform (\ref{fk}),
\be\label{compress3}
\Pi(f)(\vec{X})= \int_{\R^3} \frac{\extd^{3}k}{(2\pi)^3}\, e^{\imath\vec{k}\cdot\vec{X}}\,
\tilde{\Pi}(\tilde
f )(\vec{k})
\ee
where $\Pi$ projects into the subspace ${\CC}_{\lp/\pi}(\RR^3)$ of ordinary functions with ordinary Fourier transform having  momentum bounded by $\pi/\lp$ (so we can limit the integral in (\ref{compress3}) to $|\vec k|<\pi/\lp$ ). We also see that the image of $\DD$ is  $\hat{\CC}_{\lp}(\RR^3)$ provided one treats the boundary $|\vec k|=\pi/\lp$ appropriately. For example, the element $\zeta$ is in the image of the quantum group Fourier transform and hence of $\DD$ provided one approaches the bound from below. All of this information expressed in the diagram (\ref{SU2diag}) amounts to a `noncommutative compression theory' as explained in Section~1 and built around the extended Duflo map. 
 
Our extended Duflo  map has several further properties. First, its definition has a geometrical interpretation as an averaging 
procedure over cells of area $4\pi \lp^2$, namely we have 
\be\label{Dint}
\DD(f)(\hx) = \int_{S^2} d^2n f(\hx +\lp \vec{n})
\ee
where the integral is the normalised integral over the $2$-sphere $n^2=1$. This follows immediately from the definition and the identity \[
\int_{S^2} d^2n e^{\imath k\cdot\hx} e^{\imath\lp \vec{k}\cdot\vec{n}}
=\frac{\sin \lp |\vec{k}|}{\lp |\vec{k}|}  e^{\imath k\cdot\hx}
=\DD(e^{\imath \vec{k}\cdot \vec{X}}).\]

Second, it behaves very well on radial functions.  A first observation here
 for $f= f(r)$ any radial function in $C(\RR^3)$  is
\be\label{intrD}
\DD(f)(\hr) = \frac{1}{2\lp \hr} \int_{\hr-\lp}^{\hr +\lp} \extd u \,u f(u).
\ee
which follows directly from (\ref{intrD}) with $\hr^2= \hx_{i}\hx^{i}+\lp^2$. It in turn  implies for example that 
 \be
 \DD(\frac{1}{r}\partial_{r}(r f)) =  \frac{1}{\hr}\hat{\partial}_{r}(\hr\DD(f)),
 \ee
where 
 \[
\hat\del_r f(\hr)={f(\hr+\lp)-f(\hr-\lp)\over 2\lp}.\]
 We will see in the Section~5.4 that this see that is precisely the  radial differential for our quantum covariant calculus in polar coordinates, i.e. the latter is compatible in a nice way with the Duflo map. 
 
Next, the Duflo map on radial functions makes manifest the compression evident in the sampling theorem of Section~5.2,
\be\label{fr}
\DD(f) = \Pi(f)(\hr).
\ee
Moreover if $f$ is a radial function (an even function of $r$) the compressed map takes the same values on integers as the original function
\be\label{evfj}
\Pi(f)(\lp(j+1))= f(\lp(j+1)).
\ee
We shall prove both assertions momentarily. An immediate consequence of these two results when taken together with the sampling theorem is  \be 
 \DD(f) = \sum_{j\in \NN}(j+1) f(\lp(j+1)) \hat\delta_{j},
 \ee
for any radial function $f \in C(\RR^3)$.
As an immediate corollary we get  from the multiplication property of the $\hat \delta_{j}$ (\ref{prodel}) 
   the homomorphism property
    \be \label{prodD}
\DD(fg)= \DD(f)\cdot \DD(g).
\ee
This property is standard at the polynomial level but we see that it now holds much beyond. As another immediate corollary we have that the
action of the composition $D\equiv \phi \circ \DD$ on a radial function is given by
\be \label{theo2}
 D(f)(r)= \sum_{j\in 2\NN}(j+1) f(\lp(j+1)) \delta_{j}(r).
 \ee  
We will give an independent proof of this last statement.

We now turn to the proofs of (\ref{evfj}) and (\ref{fr}) respectively. Indeed, if $f$ is a radial function, one can evaluate $\tilde{\Pi}(\tilde f )$ thanks to the Poisson resummation formula
\be 
\tilde{\Pi}(\tilde f )(k)= 4\pi \lp^2 \sum_{j\in\NN} (j+1) f(\lp(j+1)) \frac{\sin(\lp k (j+1))}{\lp k}.
\ee
This gives 
\be 
\Pi(f)(X)= \sum_{j\in\NN} \frac{(j+1)}{|X|} f(\lp(j+1)) \left(\frac{2}{\pi}\int_{0}^\pi \extd k \,{\sin( k
(j+1))}\sin(k\frac{|X|}{\lp})\right).
\ee
If $|X|=\lp (i+1)$ the integral in the RHS is zero unless $j=i$, in which case it is $1$, this proves (\ref{evfj}).

 Let us now present the proof of (\ref{fr}). Due to the general results expressed in (\ref{SU2diag}) we can restrict to the case where 
  $f \in \CC_{\lp/\pi}(\RR^3)$;  this means that $\tilde{f}(\vec{k})$ is zero for $\lp|\vec k|\geq \pi$.
It is enough therefore to verify this property on the orbit integral of wave function with momentum $\lp|\vec k|<\pi$.
On one hand we have 
\be 
R(\DD(e^{\imath \vec{k}\cdot\vec{X}}))=\DD(R(e^{\imath \vec{k}\cdot\vec{X}}))= \DD\left(\frac{\sin |\vec k| |\vec X|}{|\vec k||\vec X|}\right),
\ee
on the other hand 
\be 
R(\DD(e^{\imath \vec{k}\cdot\vec{X}}))= \frac{\sin \lp |\vec k|}{\lp |\vec k|}R(e^{\imath k_{i}\hx^{i}}) = \frac{\sin\hr |\vec k|}{\hr |\vec k|}
\ee
where the last equality is true if  $\lp|\vec k|<\pi$ by (\ref{orbit1}). This proves (\ref{fr}) for $f\in \hat{\CC}_{\lp}(\RR^3)$.

There is also a useful direct proof of (\ref{theo2})  as follows: First, since $f$ is radial $\tilde f$ depends only on $\theta= \lp |\vec k|$.
If we  perform the integral over the angular variables we are left with
\be
\int_{0}^{\infty} \frac{\extd \theta}{2\pi^2 \lp^3} \sin^2\theta \frac{\sin (r\sin\theta)}{r\sin \theta} \frac{\theta}{\sin\theta}
\tilde f\left(\frac{\theta}{\lp}\right)
\ee
with $r\equiv |\vec{X}|/\lp$. From the fact that the integrand is even and the $\pi$ periodicity of some component of the integrand this integral can be written
\be\label{intn}
\int_{-\frac{\pi}{2}}^{+\frac{\pi}{2}} \frac{\extd \theta}{4\pi^2 \lp^3} \sin^2\theta \frac{\sin (r\sin\theta)}{r\sin \theta} 
\sum_{n\in \ZZ}\frac{\theta -n\pi}{\sin(\theta-n\pi)}\tilde f\left(\frac{\theta-n\pi}{\lp}\right).
\ee
Using the Poisson resummation formula we get 
\be 
\sum_{n\in \ZZ}\frac{\theta -n\pi}{\sin(\theta-n\pi)}\tilde f\left(\frac{\theta-n\pi}{\lp}\right)
={8\pi \lp^3}\sum_{j\in 2\NN}  (j+1)\frac{\sin(j+1)\theta}{\sin \theta} f(\lp(j+1)).
\ee
The integral (\ref{intn}) can then be written as 
\bes\nonumber
\displaystyle &&\sum_{j\in 2\NN}  (j+1) f(\lp(j+1)) \frac{2}{\pi}\int_{-\frac{\pi}{2}}^{+\frac{\pi}{2}} {\extd \theta} \sin^2\theta \frac{\sin (r\sin\theta)}{r\sin \theta} 
\frac{\sin(j+1)\theta}{\sin \theta} \\
& =&\sum_{j\in 2\NN}  (j+1) f(\lp(j+1)) \int_{SU_{2}} \extd g \, \chi_{j}(g) E_{g}(\vec{X})\\
& =&\sum_{j\in 2\NN}  (j+1) f(\lp(j+1)) \delta_{j}(r)
\ees
In the last equality we recognise  $\phi(\DD( f))$ when $\DD(f)\in \hat{\CC}_{\lp}^+(R^3)$.

 \subsection{ Differential calculus in polar coordinates}\label{rcovcal}

We have seen how a natural `radial' derivative $\hat\del_r$ appears from the radial Fourier theory and Duflo map. Let us see now how these formulae are properly defined and appear in noncommutative differential geometry. In fact it was shown recently in \cite{Ma:time} that the extra direction $\theta$ in the calculus and $\extd c$ enjoy a self-contained calculus, where $c=\hx\cdot\hx$. One has   
\[ \extd c=2\hx_i\extd \hx^i+3\imath\lp\theta\]
\[  [\extd c,\hx^i]=[\extd\hx^i,c]=3\lp^2\extd \hx^i+2\imath\lp \hx^i\theta+2\imath\lp\eps^i{}_{jk}\hx^j\extd \hx^k\]
\[  [\extd c, c]=2\lp^2\extd c+4\imath\lp(c+{3\over 4}\lp^2)\theta\]
from which it is shown that \[\extd f(c)=(\hat\del^c
f)\extd c+(\hat\del^0|_cf)\theta\] for any function $f(c)$, where $\del^c$ and $\del^0|_c$ are given in \cite{Ma:time}.   

Our first step is a convenient further change variables to $\hr=\sqrt{c+\lp^2}$ as the radial variable, which is equivalent.  Using the above formulae from \cite{Ma:time} we
compute
\[ \extd \hr=(\hd^c\hr)\extd c+(\hd^0|_c\hr)\theta={1\over 2\hr}(\extd c-\imath\lp\theta)={1\over \hr}(\hx_i\extd \hx^i+\imath\lp\theta)\]
which combined with $\extd c=\hr\extd\hr+(\extd\hr)\hr$ and
(\ref{theta}) implies\be\label{drx}
[\extd\hr,\hx^i]=[\extd\hx^i,\hr]={1\over\hr}\left(\lp^2\extd \hx^i+\imath\lp \hx^i\theta+\imath\lp\eps_{ijk}\hx^j\extd \hx^k \right)\ee
\be\label{dr}
[\extd\hr,\hr]=\imath\lp\theta,\quad
[\hr,\theta]=\imath\lp\extd\hr\ee as a  closed 2-dimensional calculus for
functions of $\hat r$ alone. Working
with such functions we define partial derivatives
\be\label{dfr}  \extd f(\hr)=(\hat\del^r f)\extd\hr+(\hat\del^0|_rf)\theta\ee
to find
\[ \hat\del^r f(\hr)={f(\hr+\lp)-f(\hr-\lp)\over 2\lp},\quad \hat\del^0|_rf(\hr)={\imath\over2\lp}\left(f(\hr+\lp)+f(\hr-\lp)-2f(\hr)\right)\]
by the same method as in \cite{Ma:time} for $\extd c$ (by solving
a recursion to compute $\extd \hr^n$), or by converting $\hat\del^c,\hat\del^0_c$ there. We see the the natural appearance
from noncommutative geometry of the finite-difference operators and 
$\hat\del^0|_r$ as precisely the finite `double derivative' in the
radial direction. By
computing $\extd f(\hat r)$ in the two bases we find also
\be\label{delir}\hat\del^i ={\hx^i \over \hr}\hat\del^r, \quad \hd_0=\hd^0|_r+\imath{\lp\over\hr}\hd^r\ee
for the change of variables between Cartesian and polar.  From these one has also
\be\label{radd0}\Del f(\hat r)=
\frac{1}{2}\left((1+\frac{\lp}{\hr})f(\hat
r+\lp)+(1-\frac{\lp}{\hr})f(\hat r-\lp)\right)\ee
as promised in Section~5.1. We see that the radial quantum differential calculus is in agreement with the Fourier transform and Duflo map  computations.

Also, for later use, we derive the braided-Leibniz rule applicable when one
function is purely `radial'. Indeed, if $f=f(\hr)$ then applying
$\extd$ to the relation $[\hx^i,f]=0$ we have
\begin{eqnarray}\label{dxf}
 {}\kern-20pt [\extd \hx^i,f]&=&[\extd f,\hx^i]=[(\hat\del^r f)\extd \hr,\hx^i]+[(\hat\del^0|_r f)\theta,\hx^i]\nonumber \\
 &=&(\del^rf)[\extd \hr,\hx^i]+(\hat\del^0|_rf)[\theta,\hx^i]\nonumber \\
 &=&({\lp^2\over\hr} \hat\del^rf-\imath\lp\hat\del^0|_r f)\extd\hx^i+
\imath{\lp\over\hr} (\hat\del^rf){\hx^i}\theta+\imath{\lp\over\hr}(\hat\del^r f)\eps_{ijk}{\hx^j}\extd  \hx^k\nonumber\\
&=&-\imath\lp(\hd_0 f)\extd\hx^i+ \imath\lp(\hat\del^if)\theta+\imath\lp\eps_{ijk}(\hat\del^jf)\extd
\hx^k.\end{eqnarray} 
using (\ref{delir}). Applying this to a computation of $\extd(\hx^if)$ we then deduce the braided-Leibniz rule
\be\label{radleib} \hd_j(\hx^i f)=\hx^i\hd_j f+\delta_{i,j}\Del f-\imath\lp\eps_{ijk}\hd_k f\ee
if $f$ is purely radial. One deduces similarly
 \be\label{waveleib} \hat\del_i(e^{\imath k\cdot\hat x}f)=e^{\imath
k\cdot\hx}\left(\imath
P_i\Del f+P_0\hat\del_if-{\lp\over\hr}(\vec
P\times\hx)_i\hat\del^rf\right)\ee
where $P_i$ are defined by (\ref{Pk}).

\subsection{Gaussian functionals}

As an application of our methods,  we now study two kinds of  quantum gaussians and their images.
We will see that these gaussians can be constructed either by applying $\phi$ term by term to a powerseries expression and computing this classical image
directly using the group Fourier theory, or working in the quantum case using   the radial sampling theorem
(\ref{sampleth}). We shall see that we can also in
principle compute the full quantum Gaussians by noncommutative calculus methods. In short, the Gaussian theory here uses
all three of the methods developed in the paper.  Finally, we use one of the Gaussians to test quantum integration.

The first Gaussian $g_\alpha(\hr)$ (say) we consider is defined as on
linear noncommutative spaces \cite{Ma:book}  by the equation
\[ \hat\del^i g_\alpha=-\alpha \hx^i g_\alpha.\]
In general this function will not be in $\CC^+_\lp(\R^3)$ but if we assume it is given by a power-series and define $\phi$ term by term, we can work with the corresponding
$G_\alpha(\vec{X})\equiv \phi(g_{\alpha})(\vec{X})$ which  should obey
\be\label{Geqn}\del^i G_\alpha=-\alpha X^i\star G_\alpha=\imath\alpha\nabla^L_i G_\alpha\ee
where   \be\label{nablaLR} \nabla_i^L E_g=
\frac{d}{dt}E_{(e^{\imath t\lp \sigma_i}g)}|_{t=0},\quad
\nabla_i^R E_g= \frac{d}{dt}E_{(ge^{\imath\lp t\sigma_i})}|_{t=0}.
\ee
are natural left and right `covariant derivative' operations. We will use both of them in Section~\ref{Fsec}  (and give a
more
algebraic description of them); at the moment we use only $\nabla^L_i$. Then equation (\ref{Geqn}) is therefore easily solved with the $SO_3$ group Fourier
transform  in terms of the $E_g$ functions. Indeed, let
\[ G_\alpha(\vec{X})=F(e^{\frac{1}{\alpha\lp^2}{\rm Tr} g})=\int \extd g\, e^{\frac{1}{2\alpha\lp^2}{\rm Tr} g}E_g(\vec{X}).\]
Applying $\del^i$ brings down a $\imath P_i$ in the integrand on the one hand,
while on the other hand $\alpha\nabla^L_i$ acts on the plane wave as a left invariant derivative on the group.
Integrating by parts we get 
\be 
\imath \alpha \nabla^L_i G_\alpha =\int \extd g (-\imath\alpha \nabla^L_i e^{\frac{1}{2\alpha\lp^2}{\rm Tr} g})E_g
=\int \extd g\, {{\rm Tr}(g\sigma_{i})\over\lp} e^{\frac{1}{2\alpha\lp^2}{\rm Tr} g}E_g
\ee
which  indeed brings down a factor  $\imath P_i$ on computing the trace. The fourier coefficients are obtained by
explicitly computing the integrals
\[
\int \extd g e^{\frac{1}{2\alpha\lp^2}{\rm Tr} g}\chi_j(g)=(-1)^{j\over 2}(j+1)
2\imath\alpha\lp^2J_{j+1}({1\over\imath\alpha\lp^2})=
(-1)^{j\over 2}d_j \delta_j({1\over\imath\alpha\lp^2})\]
for even $j$, where we recognise the Bessel function as our classicalised quantum delta-function (\ref{besselfou}). The proof is a direct computation and uses the   recurrence relation
\be\label{bessrec} J_{j+2}({1\over\imath\alpha\lp^2})+J_{j}({1\over\imath\alpha\lp^2})=2\imath\alpha\lp^2 (j+1) J_{j+1}({1\over\imath\alpha\lp^2}).\ee
Applying the character expansion in Section 5.2 we conclude that 
\[ G_\alpha=\sum_{j\in 2\NN}(-1)^{j\over 2}(j+1) \delta_j({1\over\imath\alpha\lp})\delta_j(r).\]
It is interesting to note the symmetrical role of the two delta-functions in these results suggesting an interesting
$\alpha\lp^2 \leftrightarrow1/r$ duality. Moreover, the corresponding $g_\alpha(\hr)$ is  given 
by
\[ g_\alpha=\CF(e^{\frac{1}{2\alpha\lp^2}{\rm Tr} g})=\int\extd \, g e^{\frac{1}{2\alpha\lp^2}{\rm Tr} g}e_g(\hx)\]
which is to be expected since multiplication by $\hx_j$ becomes differentiation under quantum group Fourier transform, and the part of this visible under $\phi$ is 
\be\label{galpha} g_\alpha^+=\sum_{j\in 2\NN}(-1)^{j\over 2}(j+1)  \delta_j({1\over\imath\alpha\lp^2})\hat\delta_j(\hr)\ee
This depends only on $\hr$ because the trace function is
central. The full $g_\alpha$ is given similarly by the sum over all $j\in\NN$.

 In these expressions the coefficients of $(j+1)\hat\delta_j(\hr)$ are the values in the relevant representations of  
$g_\alpha$ according to the sampling theorem of Section~5 and the general Duflo map theory. This suggests a general closed formula 
\be g_\alpha(\hr)= e^{\imath\pi{\hat\rho -1}\over 2} \delta_{\hat\rho-1}({1\over\imath\alpha\lp})\ee
provided one can make sense of the expressions on the right. 
The Bessel function here does indeed have a continuation
to complex numbers in place of $j+1$.  Indeed, if one tries to solve directly for $g_\alpha(\hr)$ using  our noncommutative differential geometry methods, their defining equation  in polar coordinates  is 
\[ \hd^r g_\alpha(\hr)=-\alpha \hr g_\alpha(\hr)\]
where the left hand side is the radial quantum differential from Section~5.4. Writing this out this becomes the recurrence relation 
\[ g_\alpha(\hr+\lp)-g_\alpha(\hr-\lp)=-2\alpha\lp \hr g_\alpha(\hr)\]
which is precisely  (\ref{bessrec}) extended to complex $j$ and formally applied to $\hr$. This also provides a   noncommutative differential geometry proof of the full version of (\ref{galpha}), where the right hand side is the part of $g_\alpha$  that it can be reconstructed from the  sampling theorem using the values on  $\hr=\lp(j+1)$. These values are required to obey 
\[ g_\alpha(\lp (j+2))-g_\alpha(\lp j)=-2\alpha\lp^2 (j+1)g_\alpha(\lp(j+1))  \]
which they are solved by Bessel functions in view of exactly (\ref{bessrec}). For (\ref{galpha}) itself we apply the $\hat\CC^+_\lp(\R^3)$ sampling theorem, which means even $j$ in (\ref{sampleth}). 

The other natural Gaussian  is defined by
\[ f_\alpha(\hx)\equiv e^{-\frac{\alpha}{2}c},\quad F_\alpha(\vec X) \equiv \sum_n \frac{(-1)^n\alpha^{n}}{2^n n!}\phi(c^n);\quad c\equiv \hx^2\]
where again we define $\phi$ as acting term by term in a powerseries expansion.  
Here we already know the Gaussian and would like to find its classical image $F_\alpha$.  If one wants to do it directly, $F_\alpha$ is characterised by
the equation
 \be
 \label{gaudiff} \frac{\partial
F_{\alpha}}{\partial \alpha}= -\frac{1}{2} X^{i}\star X_{i}\star
F_{\alpha}, \quad F_{0}(X)=1. 
\ee
{}From (\ref{Xstar}) and after
some algebra we find 
\bes\nonumber 
X_{i}\star X_{j}\star
e^{\imath\vec{P}\cdot \vec{X}} =& \left\{ (X_{i}P_{0}
+l_{p}(\vec{X}\times \vec{P})_{i})(X_{j}P_{0} - l_{p}(\vec{X}\times \vec{P})_{j})+\right. \\ 
\nonumber &\left\{\imath l_{p}(\delta_{ij} \vec{X}\cdot \vec{P}+
\epsilon_{ijk}X^{k}P_{0}+ X_{i}P_{j}-X_{j}P_{i})\right\}
e^{\imath\vec{P}\cdot \vec{X}}
 \ees 
 and hence 
 \be 
 X^{i}\star
X_{i}\star e^{i\vec{P}\cdot \vec{X}} = (\vec{X}^2
-(l_{p}\vec{X}\cdot\vec{P})^2 +3\imath l_{p}\vec{X}\cdot\vec{P} )
e^{\imath\vec{P}\cdot \vec{X}}. 
\ee 
By (as usual) thinking of
$e^{\imath \vec P\cdot \vec X}$ as a generating functional and
using $\del\over\del P_i$ to bring down powers of $X$ we find 
\be
X^{i}\star X_{i}\star F = \CS F. 
\ee 
for any polynomial $F(\vec X)$,
where
 \be 
 \CS\equiv l_{p}^2 (\vec{X}\cdot\vec{\partial})^2 +
2l_{p} \vec{X}\cdot\vec{\partial} +\vec{X}^2. 
\ee 
Assuming this
also on powerseries, we have to solve 
\be 
\frac{\partial
F_{\alpha}}{\partial \alpha}= -\frac{1}{2} \CS F_{\alpha}(X).
 \ee
 When written in polar coordinates $\CS$ depends only
on the dimensionless radial direction $\rho$, it is
explicitly given by 
\be 
l_{p}^{-2}\CS(  F) = \rho^{-1}((\rho\partial_{\rho})^2 +\rho^2 -1)\rho F), 
\ee 
The eigenfunctions of $\CS$ are again given by
Bessel functions $ \frac{J_{j+1}(\rho)}{\rho}$,  which is an
eigenvector of $\CS$ with eigenvalue $C_{j}=\lp^2(j+1)^2-\lp^2$.
 If we look for a solution which is an $L^2$ function on $\R^3$ it can
 be expanded in terms of these eigenfunctions.
 The differential equation determine the $\alpha$ dependence of the Fourier coefficients.
 The normalisation condition $F_0 =1$ can be implemented if one uses
 the identity (\ref{besselpart}). We therefore obtain
 \be
 F_{\alpha}(\vec{X}) =
e^{\frac{\alpha}{2}\lp^2}\sum_{j\in 2\NN}(j+1) e^{-\frac{\alpha\lp^2}{2}(j+1)^2} \delta_j(r).
 \ee
in the notation (\ref{besselfou}) and hence we conclude that the part of $f_\alpha$ visible under $\phi$ is
\be\label{fdelta}f_\alpha^+ = e^{\frac{\alpha}{2}\lp^2}\sum_{j\in 2\NN} (j+1)e^{-\frac{\alpha\lp^2}{2}(j+1)^2}\hat\delta_j(\hr). 
\ee
These expressions correspond to the quantum group Fourier transform of  
\be
\tilde f_{\alpha}^+(g) =e^{\frac{\alpha}{2}\lp^2}\sum_{j\in 2 \NN} (j+1)e^{-\frac{\alpha\lp^2}{2}(j+1)^2} \chi_{j}(g).
\ee
This can be re-expressed via the Poisson formula 
as 
\be
\tilde f_{\alpha}^+(g) =\frac{\sqrt{2\pi}}{4 \lp^3\alpha^{3/2}} \sum_{n \in \ZZ} 
\frac{(\theta -\pi n )}{\sin(\theta -\pi n )} e^{-\frac{1}{2 \alpha \lp^2}(\theta -\pi n )^2},
\ee
with $\mathrm{Tr}(g)=2 \cos \theta$.
We can Fourier transform back this expression and obtain after simple algebra
\be 
F_{\alpha}(r) =\frac{\sqrt{2\pi}}{\pi\lp^3 \alpha^{3/2}}    \int_{0}^{+\infty}{\extd \theta}\,
\theta \frac{\sin(\rho \sin \theta) }{\rho}e^{-\frac{1}{2\alpha \lp^2}\theta^2}
\ee
which can be expressed in terms of the dimensionful  variables as a three dimensional integral
\be
F_{\alpha}(\vec{X})= \left(\frac{2\pi}{\alpha}\right)^{3/2} \int \frac{d^3 k}{(2\pi)^3}
\frac{\sin \lp |\vec{k}|}{\lp |\vec{k}|}e^{\imath \frac{\sin \lp |\vec{k}|}{\lp |\vec{k}|}  \vec{k}\cdot \vec{X}} e^{-\frac{1}{2\alpha\lp^2}k^2},
\ee
from which it is clear to see that we recover the usual expression for the Gaussian function in the limit where $\lp \to 0$.

For this Gaussian we can also derive the full expansions immediately from the sampling theorem; the right hand side of 
(\ref{fdelta}) is the part in $\CC^+_\lp(\R^3)$ that can be reconstructed from (\ref{sampleth}) using only even $j$, with the values
given here by $\hr=\lp(j+1)$ inserted into $f_\alpha(\hr)=e^{-{\alpha\over 2}(\hr^2-\lp^2)}$. We then apply $\phi$ for
the expansion of $F_\alpha(\vec X)$. The full $f_\alpha$ is similarly given by the same expansion (\ref{fdelta}) but now with all $j\in \NN$.

It is interesting to look at $f_\alpha$ using our radial quantum differential
calculus.  Clearly from (\ref{delir}) we have  
\[
\hat\del^i f_\alpha(\hr)=-{\sinh(\alpha\lp{\hat r})\over \lp{\hat r}}\hat x^if_\alpha(\hr)e^{-\frac{\alpha}{2}\lp^2}
\]
 as the noncommutative differential equation obeyed by $f_\alpha$. 
 If one could apply $\phi$ to the right hand side  (i.e. make the required $\star$-product) we should likewise obtain $\del^iF(X)$ according to the results above. This would be rather hard to do directly, however. Working in polar coordinates, we also have from Section~5.4 that 
 \[\hat\del^r f_\alpha=-\sinh(\alpha\lp\hr)e^{-{\alpha\over 2}\lp^2}f_\alpha.\]
 \be\label{delrf} \Del f_\alpha=\left(\cosh(\alpha\lp{\hat r})-{\lp\over\hr}\sinh(\alpha\lp\hr)\right) e^{-\frac{\alpha}{2}\lp^2} f_\alpha.\ee

As an application of Gaussians let us consider physically reasonable algebra $\hat C_\lp^{\rm Gaussian}(\R^3)$ of 
`Gaussian wave functions' spanned by functions of the form $f_\alpha e_g$, with $\alpha\ge 0$. As shown in the Duflo 
map section this algebra is in fact not
much bigger than the observable subalgebra $\hat\CC^+_\lp(\R^3)$ (which corresponds to Gaussian wave functions with
$\alpha=0$). Moreover, all its elements are integrable for $\int$ and $\int_+$ in Section~4.1 due to the rapid decay at large
spin. This suggests that the quantum integration method continues to be translation-invariant, in the
sense
 \[
  \int_+ \hat\del_i(e^{\imath k\cdot\hx}f_\alpha)=0\]
  and similarly with $\int$.  In view of  (\ref{waveleib}) such integrals will again be absolutely convergent due to the $f_\alpha$ factor, hence
allowing differentiation ${\del\over\del k_m}|_{k=0}$ inside the integral. Hence although not quite the same,  
translation invariance carries similar information to translation-invariance for products of polynomials with
$f_\alpha$. This implies that 
 \be\label{intf}\int_+ \hat\del_i(\hx^if_\alpha)=0\ee
 where we sum over $i$ (this can be computed more easily as the integrand remains in the centre of the algebra). Such translation invariance in noncommutative geometry looks innocent enough it contains a lot of information and can be  a useful check on our work. Thus, we use our noncommutative polar coordinates to compute the $\hat\del_i$ and use the results of Section~5.4 to find
 \begin{eqnarray*} \int_+ \hd_i(\hx^if_\alpha)
&=&\sum_{j\in 2\NN}d_je^{-{\alpha\over 2}\lp^2 d_j^2}\left(3 d_j
\cosh(\alpha\lp^2 d_j)-(d_j^2+2)\sinh(\alpha\lp^2 d_j)\right)\end{eqnarray*}
using (\ref{delrf}), (\ref{radleib}) and $\hr=\lp d_j=\lp(j+1)$ in the relevant representation.  The claim is that this is necessarily zero since the left hand side is a noncommutative total divergence, and similarly for $\int$ with the series summed over $j\in\NN$. It is possible to verify this directly, as an independent check of many of our formulae.

\section{effective action}

{}From our above analysis relating classical fields to noncommutative geometry and the Duflo map theory 
we have made precise the idea that the available space of noncommutative fields $\hat C_\lp(\R^3)$ that results from integration over 2+1 quantum gravity is a space of 
bounded momenta which cannot resolve spacetime geometry at smaller scales than Planck length. We have seen that this is not much bigger than the `physical subspace' $\hat\CC_\lp^+(\R^3)$ which has classical counterpart $\CC^+_\lp(\R^3)$ of fields with manifestly bounded momenta. 
Moreover the sampling theorem shows how it is possible to have a spacetime which is at the same time continuous and
discrete. That is the physical fields are function on $\RR^3$ but equivalently can be view as function on a discrete
spacetime lattice (at least for radial fields), this illustrates nicely how one can implement at the same time 
a physical cut-off while preserving the action of a continuous symmetry group (The Euclidean group here).

We can now combine these different ingredients in a comparison of
the noncommutative actions for the effective field theory.
Firstly, the one proposed in \cite{BatMa} coming out of the
noncommutative differential calculus is as follows. The canonical
extension to differential forms on a quantum group in this case
becomes that $\extd\hat x^i,\theta$ mutually anticommute as usual.
We also define the Hodge $*$-operator as usual, say with Euclidean
signature in the 4-dimensional cotangent space. (Even though the
base space is 3-dimensional the noncommutative geometry at each
point is 4-dimensional.) We define the integral of a 4-form as the
$\int$ above of the coefficient of $\theta\extd\hat
x^1\cdots\extd\hat x^3$ (say). If $f\in \hat{C}_\lp(\R^3)$ is a
real $f^\dagger=f$ noncommutative field then the action is (cf.\cite{BatMa}) \begin{eqnarray}
\label{qgaction}
S&=&\int_+ \extd   f{}^*\extd f=\int_+\extd(  f{}^*\extd f)-\int  f\extd *\extd f=-\int_+   f\hat\del_\mu\hat\del^\mu f\nonumber
=-\int_+   f(\hat\nabla^2+c^{-2}\hat\del_0^2) f\\
& \approx& -\int_\phi   f(\hat\nabla^2+c^{-2}\hat\del_0^2) f
=-\int_{\R^3}\phi( 
f)\star(\nabla^2+c^{-2}\del_0^2)\phi(f)\extd^3X \end{eqnarray} where  the total derivative vanishes due to
$\int$ being translation-invariant on $\hat{C}_\lp(\R^3)$. The
manipulations with the Hodge $*$ are the same as in the
commutative case. The index $\mu=0,1,2,3$ and is raised using
whatever signature (constant) metric is being used in the Hodge
$*$-operator -- we have chosen (say) the Euclidean metric with
scale $c$ in the $\theta$ direction. At the end we use the map
$\phi$ and $\del_0$ the classical counterpart of (\ref{del0}) with
respect to this.

This action is essentially the same as the one obtained  from 2+1 quantum
gravity as the effective theory in \cite{EL} when the scale parameter $c\to\infty$.
There \be S_{\rm eff}=-\int_{\R^3} \phi\star\nabla^2\phi\
\extd^3X=\int_{\R^3} \del^i \phi\star\del_i\phi\ \extd^3X\ee for
classical fields $\phi$. The second equality
here follows easily by Fourier transform as noted already in
\cite{EL}. Finite $c$ here does not appear to have any role in the
physical theory (and indeed we shall argue that this `extra dimension' is not some kind
of external time). But there is an extra direction $\hat\del_0$
which is forced by the noncommutative geometry and which does suggest an intrinsically 4-dimensional picture of some kind. Dropping it
would render the calculus (\ref{calc}) nonassociative, but we have seen many places
where this $\hat\del_0$ does play a critical role if not in the action itself.  It is worth noting 
that from a mathematical point of view the choice $c=1$ in
(\ref{qgaction}) suggests a slightly different action \be 
S'_{\rm eff}= {2\imath\over\lp}\int_{\R^3}  \phi\star \del_0 \phi\
\extd^3X \ee which might be of interest in some other context. 
Aside from requiring $c\to \infty$ we see that the approach \cite{BatMa} `working
up' from noncommutative geometry and the approach \cite{EL}
`working down' from the Ponzano-Regge model agree.

Next, while the 2+1 quantum gravity $\star$ is tied to $SO_3$ our work suggest a natural extension to  $SU_2$
 which would allow half-integer spins in the underlying spin network. 
As we have seen the fuller space of physical fields is obtained as a doubling of the 
space of $SO_{3}$ fields which respect the star product structure (adding the generator $\zeta$). 
One therefore expects the effective theory 
to be a `complexification' of the $SO_{3}$ theory just described.
We have also seen that one can equivalently 
describe the $SU_{2}$ theory in an extended  $\tilde\star$-product form by indeed introducing an 
 extra variable $T$ and with physical modes recovered as solutions to a laplace equation. Moreover, the scalar product can be given a $4$-dimensional perspective.  We have also seen  in Section~5.2 that this extra variable $T$ it is related to the
passage from $SU_2$ to $SO_3$ i.e. from a `finer theory' to a coarser one. In other words $e^{\imath T}$ is related to a `coarse-graining'
step. In the finer theory we would use the integral $\int$ summing over all characters not only the even ones, have the
odd $\hat\delta_j(\hr)$ and have a twice as good sampling
and power series approximations according to the results of Section~5. We conclude that the meaning of $T$ is a not `time' but a renormalisation group flow parameter and that this is becoming intermeshed
with the effective 3-dimensional noncommutative geometry in the model. 
It therefore suggest that that the effective field theory could be given  a 4-dimensional perspective with the extra `time' 
equation having the interpretation of a renormalisation group equation; an eventuality recently raised in \cite{GL}. 

The appearance of an `extra dimension' in the noncommutative geometry is, moreover, a typical feature as a kind of anomaly that arises when one wishes to preserve a quantum symmetry\cite{Ma:time} and we can speculate that the renormalisation group should enter in some generality in connection with this. For example, the 4-dimensional bicrossproduct noncommutative spacetime, a different model to the one in the present paper, has standard 4-dimensional differential calculus, but is it not covariant under the relevant (bicrossproduct) Poincar\'e quantum group, one needs a 5-dimensional calculus \cite{Sit}. Likewise all enveloping algebras of simple Lie algebras viewed as a noncommutative spacetimes incur such an anomaly for their Ad-invariant differential calculus \cite{BegMa}. Therefore the analysis in the present paper for our specific model appears indicative of a quite general phenomenon.

\section{Twist operator}\label{Fsec}

Although any $\star$-product by construction can be viewed as the
action of a bidifferential operator, one has much more information
if one can express this as the action of an invertible cochain
twist  element $F\in H\tens H$ where $H$ is some symmetry algebra
acting covariantly on the undeformed algebra and \be\label{Fp}
\star= m F.\ee Here $m$ denotes the undeformed product, in our
case $m: C(\RR^3)\otimes C(\RR^3) \to C(\RR^3)$ is the pointwise
product. Evidence that $F$ should exist for the usual
$\star$-product for $U(su_2)$ (for example) and $H$ the enveloping
algebra of a certain Lie algebra is in \cite{BegMa}, where $F$ is
given to the lowest two orders.  We ask if $F$ similarly exists
for the quantum gravity $\star$-product (\ref{star}).

 What appears to come out naturally in this  case is not exactly such an $F$ but something a little weaker, but which we find in closed form. We find an invertible $F\in H\tens H$ obeying (\ref{Fp}) but do not require $H$ to be a Hopf algebra. Rather, we take for $H$  the usual Heisenberg algebra of multiplication and differentiation on $\R^3$, which acts on $C(\R^3)$ as a vector space (the usual Schroedinger representation) but does not have a coproduct so there is no meaningful sense in which it can act covariantly on the algebra of $C(\R^3)$.  This is therefore in the same spirit as the Moyal product but not exactly in the setting needed for quantum group methods.

Let $\hat X^i$ (which is $X^i$ acting by multiplication) and
$\hat{P}_i=-\imath\del_i$ be the usual Heisenberg algebra
generators. Let $\hat{J}_i\equiv \lp \epsilon_{ijk}\hX^j\hp^k$ be
the usual orbital angular momentum scaled as a realisation of
$\hat{C}_\lp(\R^3)$ and let \be \hp_0 = \sqrt{1-\lp^2
\hp^2}=\der.\ee
 We consider two operators in the Heisenberg algebra,
\be \nabla_i^L \equiv \imath(\hX_i \hp_0 + \hat J_i)=\imath
X_i\star(\ ),\quad \nabla_i^R \equiv \imath(\hX_i \hp_0 - \hat
J_i)=(\ )\star \imath X_i, \ee in terms of (\ref{Xstar}) verified
on plane waves. Since the $\star$-product is associative and obeys
the relations (\ref{su2}) it follows immediately that \be
[\nabla_i^L,\nabla_j^L]=-2\lp\epsilon_{ij}{}^k\nabla_k^L,\quad
[\nabla_i^R,\nabla_j^R]=2\lp\epsilon_{ij}{}^k\nabla_k^R,\quad
[\nabla_i^R,\nabla_j^L]=0. \ee
 Left and right $\star$-multiplication corresponds under Fourier transform (\ref{fou}) to differential operators on $C(SU_2)$ of left and right multiplication in the group. Hence (or by direct computation) their action on
plane waves $E_g$ exponentiates to \be
e^{\vec k_2\cdot\nabla^L}E_{g_1}=\phi(e^{\imath \vec k_2\cdot \hat
x^i})\star E_{g_1}=E_{g_2}\star E_{g_1}=E_{g_2g_1}\ee if $g_2 =
e^{\imath  \lp k_2^i\sigma_i}$. Similarly for $e^{\vec
k_2\cdot\nabla^R}$. The operators here are the same ones as introduced geometrically
in (\ref{nablaLR}). 

Let us also introduce the operator $\hat{k}_i$ such that \be
\frac{\sin \lp |\hat{k}|}{\lp |\hat{k}|} \hat{k}_i =\hat{P}_i \ee
that is \be \hat{k}_i = \frac{\arcsin \lp |\hp|}{\lp| \hp|} \hp_i
= \hp_i(1 + \lp^2 \frac{|\hp|^2}{6}+\cdots), \ee where the RHS is
understood as a perturbative expansion in $\lp$.

We are now ready to define \be F=:e^{-\imath\hat X^i\tens \hat
P_i}e^{\nabla^{Ri}\tens\hat k_i}:\ee where the normal ordering is
to put all the $\hat k_i$ and $\hat P_i$ operators to the right.
Then \be
 F(E_{g_1}\otimes E_{g_2} )
 = \left(e^{- \imath\vec P_2\cdot \hat X\tens 1} e^{\vec k_2\cdot\nabla^R\tens 1}\right)(E_{g_1}\otimes E_{g_2})
=(E_{g_2^{-1}})E_{g_1g_2}\otimes E_{g_2} \ee obeys (\ref{Fp}) as
required since the undeformed product is commutative. There is an
inverse \be F^{-1}=:e^{-\nabla^{Ri}\tens\hat k_i}e^{\imath\hat
X^i\tens \hat P_i}: \ee by a similar computation. Also note that
there is a similar operator \be F_L=:e^{-\imath \hat P_i\tens\hat
X^i}e^{\hat k_i\tens\nabla^{Li}}:\ee with ordering of the $\hat
P_i,\hat k_i$ to the right that does the same job in providing an
equally good twist operator.  The two do not commute. Finally,
these twist operators all begin with $1\tens 1$ in an
$\lp$-expansion as in the cochain twist theory. One finds for
example,
\[ F=1\tens 1-\lp\imath\eps_{ijk}\hat X^i\hat P^j\tens \hat P^k-\imath \frac{\lp^2}{2}\hat X^i\hat P^2\tens \hat P_i+ \imath\frac{\lp^2}{6}\hat X^i\tens \hat P_i\hat P^2+O(\lp^3)\]
 The first term lives in the tensor square of the Lie algebra generated by $\hat J_i,\hat P_i$, which agrees with the general cochain proposed in \cite{BegMa} at this order. The next terms can be written in terms of
 \be \del_0=-\imath\frac{\lp}{2}\hat P^2+O(\lp^2)\ee
 which exhibits its role in the twist operator and takes our $F$ out of the cochain setting.

We can similarly express the operator \be\label{R} R= :e^{ 2\imath
\hat J^i\otimes\hat{k}_i}:=F^{-1}F_{L21} \ee which reproduces the
quantum double braiding \be R(E_{g_1}\otimes
E_{g_2})=E_{g_2g_1g_2^{-1}}\otimes E_{g_2} \ee known from the
action of $D(U(su_2))$ on the deformed algebra, as for the double of any group algebra\cite{Ma:book}. As for the action
of the quantum double of any Hopf algebra, this necessarily obeys
\be \star (R-\sigma)=0, \ee where $\sigma$ is the flip operator
trivially interchanging the two copies, and the Yang-Baxter
equation \be R_{12}R_{13}R_{23}=R_{23}R_{13}R_{12}, \ee where the
numerical suffices denote the position on a 3-fold tensor product.
That $R$ is also given by the second expression in (\ref{R})
follows by evaluation on plane waves $E_{g_1}\tens E_{g_2}$ and
reminds us of the theory of cochain twists. Since $R$ is on any
quantum double is definitely not triangular one could not expect
$R=F^{-1}F_{21}$ -- instead we see the role of  the second twist
$F_L$.


\begin{thebibliography}{99}
\bibitem{EL}
  L.~Freidel and E.~R.~Livine,
\newblock Ponzano-Regge model revisited. III: Feynman diagrams and effective field theory,
\newblock hep-th/0502106.


\bibitem{BatMa} E. Batista and S. Majid.
\newblock Noncommutative geometry of angular momentum space $U(su_2)$.
\newblock {\em J. Math. Phys.} 44 (2003) 107-137.

\bibitem{Wit}
E. Witten.
\newblock 2+1 Gravity as an Exactly Soluble System, {\em Nucl. Phys. B} 311:46--78, 1988.


\bibitem{TurVir}
V. G. Turaev and O. Y. Viro.
\newblock State sum invariants of 3-manifolds and quantum 6j-symbols {\em Topology} 31(1992)865-902. 

\bibitem{PR0}
G Ponzano, T Regge.
\newblock Semi-classical limit of Racah coefficients.
\newblock In {\em Spectroscopic and group theoretical methods in physics}, ed. Bloch, North Holland (1968)

\bibitem{tHooft}
  G.~'t Hooft.
  \newblock Quantization of point particles in 2+1 dimensional gravity and space-time
  discreteness.
  \newblock  Class.\ Quant.\ Grav.\  {\bf 13}, 1023 (1996)
  [arXiv:gr-qc/9601014].

\bibitem{BaiMul}
F.A. Bais and N.M. M\"{u}ller.
\newblock Topological field theory and the
quantum double of $SU(2)$.
\newblock {\em Nucl. Phys. B}530:349--400, 1998.

 \bibitem{Sch}
B.J. Schroers.
\newblock Combinatorial quantization of Euclidean gravity
in three dimensions.
\newblock in {\em Quantization of singular symplectic quotients}, eds N. P. Landsman, M. Pflaum and M. Schlichenmaier, Progress in Mathematics 198, pp 307--328. Birkhauser 2001.


 \bibitem{AmeMa}
 G. Amelino-Camelia and S. Majid.
 \newblock Waves on noncommutative spacetime and gamma-ray bursts.
 \newblock {\em Int. J. Mod. Phys.} A15:4301--4323, 2000.

 \bibitem{Ma:time}
S. Majid.
\newblock Noncommutative model with spontaneous time generation and Planckian bound.
 \newblock {\em J. Math. Phys.} 46:103520,  2005, 18pp.

\bibitem{sampling}
J.J. Benedetto, P.J.S.G. Ferreira.
{\em Modern Sampling theory}, Birkhauser (2001).

\bibitem{Kempf}
  A.~Kempf. 
  \newblock Fields over unsharp coordinates. 
  Phys.\ Rev.\ Lett.\  {\bf 85}, 2873 (2000), hep-th/9905114.


 \bibitem{BegMa}
E.J. Beggs and S. Majid.
\newblock Quantization  by cochain twists and nonassociative differentials.
\newblock  Math.QA/0506450.

\bibitem{ARS}
A.Yu. Alekseev, A. Recknagel, V. Schomerus.
\newblock Noncommutative
world-volume geometries: branes on $SU(2)$ and fuzzy spheres.
\newblock {\em J.
High Energy Phys.} 9909:023, 1999.



\bibitem{Gut}
S. Gutt.
\newblock An explicit $*$ product on the cotangent bundle of a Lie group.
\newblock Lett. in Math . Phys. 7:249--258, 1983.

\bibitem{KL}
  L.~Freidel and K.~Krasnov,
 \newblock The fuzzy sphere *-product and spin networks. 
 \newblock J.\ Math.\ Phys.\  {\bf 43}, 1737 (2002), hep-th/0103070.
 

\bibitem{Wor}
S.L. Woronowicz, Com. Math. Phys. {\bf 111}, 613 (1987).


\bibitem{Ma:book}
S. Majid.
\newblock {\em Foundations of quantum group theory}. Paperback edn. C.U.P. 2000.

\bibitem{KemMa}
  A.~Kempf and S.~Majid,
\newblock Algebraic q integration and Fourier theory on quantum and braided spaces.
  \newblock J.\ Math.\ Phys.\  {\bf 35}, 6802 (1994), hep-th/9402037.

\bibitem{Ma:twi}
S. Majid, 
\newblock Noncommutative Differential Geometry and Twisting of Quantum Groups.
\newblock In L.M.S. Lect. Notes {\bf 290} 175-190 (2001).



\bibitem{PRI}
  L.~Freidel and D.~Louapre.
\newblock Ponzano-Regge model revisited. I: Gauge fixing, observables and interacting spinning particles. 
\newblock  Class.\ Quant.\ Grav.\  {\bf 21}, 5685 (2004), hep-th/0401076.

 \bibitem{GL}
F. Girelli and E. Livine,
Physics of deformed special relativity: Relativity principle revisited,
gr-qc/0412004.


\bibitem{Sit}
A. Sitarz.
\newblock Noncommutative differential calculus on the kappa-Minkowski space.
\newblock Phys.Lett. {\bf B349} (1995) 42-48.


 \end{thebibliography}
\end{document}